# Analytic Adjoint Solutions for the 2D Incompressible Euler Equations Using the Green's Function Approach

**Carlos Lozano, Jorge Ponsin**

**Computational Aerodynamics**

**National Institute of Aerospace Technology (INTA)**

## Abstract

The Green's function approach of Giles and Pierce is used to build the lift and drag based analytic adjoint solutions for the two-dimensional incompressible Euler equations around irrotational base flows. The drag-based adjoint solution turns out to have a very simple closed form in terms of the flow variables and is smooth throughout the flow domain, while the lift-based solution is singular at rear stagnation points and sharp trailing edges owing to the Kutta condition. This singularity is propagated to the whole dividing streamline (which includes the incoming stagnation streamline and the wall) upstream of the rear singularity (trailing edge or rear stagnation point) by the sensitivity of the Kutta condition to changes in the stagnation pressure.

## 1. Introduction

The aim of this paper is to apply the program developed in (Giles & Pierce, 1997; Giles & Pierce, 1998; Giles & Pierce, 2001) to the 2D incompressible adjoint Euler equations in order to obtain the analytic adjoint solutions corresponding to aerodynamic lift and drag. The idea of the method is based upon the observation that, with appropriate boundary conditions for the linearized and adjoint problems, the adjoint variables at a particular point correspond to the functional of interest (typically, though not exclusively, aerodynamic lift or drag) evaluated using the Green's function for that point (Giles & Pierce, 2000). In the analysis in (Giles & Pierce, 1997; Giles & Pierce, 2001) for the quasi-1D and 2D Euler equations, the approach in each case was to construct $d$ linearly independent source vectors $f^{(j)}(\vec{x}_0)$ which produced Green's functions $\delta U^{(j)}(\vec{x},\vec{x}_0)$ of a simple form as solutions to the equations

$$L\delta U^{(j)}(\vec{x},\vec{x}_0) = f^{(j)}(\vec{x}_0)\delta(\vec{x}-\vec{x}_0), \quad j=1,\ldots,d, \tag{1}$$

where $L$ are the linearized flow equations, $\delta(\vec{x}-\vec{x}_0)$ is the Dirac delta function and $d$ is the number of flow equations, 3 for quasi-1D flow and 4 for 2D compressible inviscid flow, respectively. If $I^{(j)}(\vec{x}_0)$ is the value of the linear functional evaluated with $\delta U^{(j)}(\vec{x},\vec{x}_0)$ and $\psi$ is the adjoint state associated to that functional then, by definition, and provided that suitable boundary conditions hold,

$$I^{(j)}(\vec{x}_0) = \psi^T f^{(j)}(\vec{x}_0) \tag{2}$$

and hence the adjoint variables can be computed from $I^{(j)}(\vec{x}_0)$ by inverting Eq. (2).

The above program allows to gain insight into the nature of the Green's function and the adjoint solution, making it possible in particular to locate singularities and discontinuities in the adjoint variables. Furthermore, given a numerical adjoint solution and a set of source vectors $f^{(j)}$, the corresponding linear functionals $I^{(j)}$ can be evaluated using Equation (2), providing a means of verifying a number of the adjoint solution properties.

The adjoint equations were introduced for design optimization in the field of computational aerodynamics by Jameson (Jameson, 1988), and have been since extended to a variety of applications such as error estimation and mesh adaptation (Venditti & Darmofal, 2002) and stability analysis (Luchini & Bottaro, 2014), among many others. From the computational viewpoint, adjoint methods can be devised in two ways (Nadarajah & Jameson, 2000), which differ in how the discretized adjoint equations are obtained. In the continuous approach, one discretizes the adjoint p.d.e., while in the discrete approach the adjoint equations are obtained directly from the discretized flow equations. An important step in the application of adjoint methods is the development and verification of adjoint codes. This task is complicated by the lack of benchmark test cases, including exact solutions, and verification is usually done indirectly by comparing sensitivities with finite differences, which may lead to erroneous conclusions (Lozano & Ponsin, 2012), or (in the case of discrete adjoint solvers) by cross-checking with the linearized solver (Giles, et al., 2003).

From a very early stage, and starting with Jameson’s groundbreaking work, developments in adjoint methods have traditionally focused on the application of the method to the optimal design (Peter & Dwight, 2010) of aircrafts (Kenway, et al., 2019), ships (Kroger, et al., 2018) and automobiles (Othmer, 2014), or to goal-oriented mesh adaptation (Fidkowski & Darmofal, 2011). Comparatively less effort has been put on analyzing the properties of the adjoint solutions, with some notable exceptions starting with the work of Giles and Pierce (Giles & Pierce, 1997) (Giles & Pierce, 1998) (Giles & Pierce, 2001) (see also (Baeza, et al., 2009) (Marta, et al., 2013) (Lozano, 2018) (Lozano, 2019) (Peter, et al., 2022)). A few exact solutions to the adjoint equations are known. For quasi-1D inviscid flows, the Green’s function approach has provided a tool to generate exact adjoint solutions (Giles & Pierce, 2001) (Lozano, 2018) in terms of the base flow solution. In inviscid 2D/3D flows, the entropy variables provide another exact solution for an output measuring net entropy flux across boundaries (Fidkowski & Roe, 2010) (Lozano, 2019). A closely related solution, corresponding to a nearfield computation of aerodynamic drag, has been recently discovered by the authors (Lozano & Ponsin, 2021). Finally, an exact solution for the adjoint Navier-Stokes equations corresponding to Blasius laminar boundary layer has been derived in (Kühl, et al., 2021).

In this paper, we present a closed-form adjoint solution for the two-dimensional incompressible Euler equations for irrotational base flows based on the Green’s function approach. This solution provides a cheap adjoint field for design or adaptation purposes that, perhaps more importantly in practical terms, can also serve for verification and debugging of adjoint solvers. But they can also serve to disentangle the properties of the solutions to the adjoint equations, which, despite being linear partial differential equations, give rise to a rich zoo of structures comprising singularities along stagnation streamlines, stagnation points and sharp trailing edges (Giles & Pierce, 1997) (Lozano, 2019) (Giles & Pierce, 2002) (Giles & Pierce, 1999), and also to confirm the existence of an adjoint singularity at solid walls suggested by recent results (Peter, et al., 2022) that could explain the numerical divergence first observed in (Lozano, 2019).

## 2. 2D incompressible adjoint Euler equations

In order to be as self-contained as possible, and to fix notations, we begin by recalling a few facts regarding the adjoint equations. In the present paper, we will concentrate on the incompressible Euler equations and consider linearizations around an irrotational base flow. It is well known that in such case the base flow can be derived either from a potential or a stream function that obey Laplace's equation. We could consider perturbations to this Laplace equation which would lead to an adjoint solution obeying the Laplace equation as well (as in e.g. (Giles & Pierce, 1999)). We will not follow this approach here. Instead, we will consider perturbations to the full incompressible Euler equations and obtain the corresponding full incompressible adjoint Euler equations, and specialize later to perturbations around an irrotational base flow.

We will focus, for definiteness, on steady, two-dimensional, incompressible, inviscid flow on a domain $\Omega$ with far-field boundary $S^{\infty}$ and wall boundary $S$ (typically an airfoil profile). The flow is governed by the incompressible Euler equations

$$R(U) = \nabla \cdot \vec{F}(U) = 0, \tag{3}$$

where

$$U = \begin{pmatrix} p \\ \rho u \\ \rho v \end{pmatrix}, \quad \vec{F} = \begin{pmatrix} \rho \vec{v} \\ \rho \vec{v} u + p\hat{x} \\ \rho \vec{v} v + p\hat{y} \end{pmatrix}. \tag{4}$$

Here, $\rho$ is the (constant) density, $\vec{v} = (u,v) = (q\cos\phi, q\sin\phi)$ is the velocity (where $q^2 = u^2 + v^2$ and $\phi$ is the local flow angle), $p$ is the pressure and $\hat{x}, \hat{y}$ are the Cartesian unit vectors. The flow obeys a non-transpiration boundary condition $\vec{v} \cdot \hat{n} = 0$ on $S$ (where $\hat{n}$ is the outward-pointing unit normal vector at the boundary) and it is assumed to approach the far-field state $(q_{\infty}\cos\alpha, q_{\infty}\sin\alpha)$ at the far-field $S^{\infty}$, $\alpha$ being the angle of attack.

In order to set-up the adjoint problem, an objective function needs to be defined, which is taken to be the lift or drag coefficient

$$I = \int_S C_p (\hat{n} \cdot \vec{d}) ds, \tag{5}$$

where $\vec{d} = \vec{d}_D = (\cos\alpha, \sin\alpha)$ for drag and $\vec{d} = \vec{d}_L = (-\sin\alpha, \cos\alpha)$ for lift, $C_p = c_\infty^{-1}(p - p_\infty)$ is the non-dimensional pressure coefficient, $p_\infty$ is the pressure of the free-stream state at the far-field and $c_\infty = \rho_\infty q_\infty^2 \ell / 2$ is a normalization factor ($\ell$ is a reference length). We also need to consider linear perturbations to (3) and (5). For definiteness, we will focus on design applications, for which perturbations in the flow solution, $\delta U = (\delta p, \rho\delta u, \rho\delta v)^T$, arise as a result of changes, $\delta\vec{x}$, in the geometry of the boundary *S*. Linearizing (3) with respect to $\delta U$ yields

$$\nabla \cdot (\vec{A}\delta U) = f, \tag{6}$$

where

$$\vec{A} = \frac{\partial}{\partial(p, \rho u, \rho v)} \begin{pmatrix} \rho\vec{v} \\ \rho u\vec{v} + p\hat{x} \\ \rho v\vec{v} + p\hat{y} \end{pmatrix} = \begin{pmatrix} 0 & \hat{x} & \hat{y} \\ \hat{x} & \vec{v} + u\hat{x} & u\hat{y} \\ \hat{y} & v\hat{x} & \vec{v} + v\hat{y} \end{pmatrix} \tag{7}$$

are the flux Jacobians and the residual *f* is typically zero for shape optimization problems. $\delta U$ obeys the following boundary condition at solid walls:

$$\delta\vec{v} \cdot \hat{n} = w \ \text{ on } S, \tag{8}$$

where $w = -\vec{v} \cdot \delta\hat{n} - [(\delta\vec{x} \cdot \nabla)\vec{v}] \cdot \hat{n}$, with $\delta\hat{n}$ being the variation in the boundary normal induced by the deformation of the geometry. In the far-field, $\delta U$ is chosen such that there is no perturbation to the flow far-field boundary conditions, unless one is specifically considering the effect of perturbations in far-field parameters such as angle of attack.

The variation of the cost function (5) with respect to perturbations in the flow and the geometry is (Castro, et al., September 2007; Anderson & Venkatakrishnan, 1999; Delfour & Zolésio, 2011),

$$\delta I = \int_S c_\infty^{-1}(\vec{d} \cdot \hat{n})\delta p\, ds + \int_S c_\infty^{-1}(\delta\vec{x} \cdot \hat{n})(\vec{d} \cdot \nabla p)\, ds. \tag{9}$$

where $\delta p$ reflects the variation of the solution in response to the changing geometry.

Eq. (9) can be efficiently computed with the adjoint approach (Jameson, 1988), where adjoint variables $\psi = (\psi_1, \vec{\varphi})^T$, with $\vec{\varphi} = (\psi_2, \psi_3)$, are introduced over the entire flow domain, and the objective function *I* is reformulated as

$$L = I - \int_\Omega \psi^T R(U)\, d\Omega. \tag{10}$$

Linearizing the Lagrangian (10) with respect to flow and geometry perturbations gives

$$\delta L = \int_S c_\infty^{-1}(\vec{d} \cdot \hat{n})\delta p\, ds + \int_S c_\infty^{-1}(\delta\vec{x} \cdot \hat{n})(\vec{d} \cdot \nabla p)\, ds - \int_\Omega \psi^T \nabla \cdot (\vec{A}\delta U)\, d\Omega + \int_\Omega \psi^T f\, d\Omega. \tag{11}$$

After integration by parts and rearrangement, this yields (Castro, et al., September 2007; Anderson & Venkatakrishnan, 1999):

$$
\begin{aligned}
&\delta L = \int_S c_\infty^{-1}(\vec{d}\cdot\hat{n})\delta p ds + \int_S c_\infty^{-1}(\delta\vec{x}\cdot\hat{n})(\vec{d}\cdot\nabla p)ds + \int_\Omega \nabla\psi^T\cdot\vec{A}\delta U d\Omega \\
&-\int_S \psi^T\hat{n}\cdot\vec{A}\delta U ds - \int_{S_\infty}\psi^T\hat{n}\cdot\vec{A}\delta U ds + \int_\Omega \psi^T f d\Omega = \\
&\int_\Omega \nabla\psi^T\cdot\vec{A}\delta U d\Omega + \int_S (c_\infty^{-1}(\vec{d}\cdot\hat{n}) - \hat{n}\cdot\vec{\varphi})\delta p ds - \int_{S_\infty}\psi^T\hat{n}\cdot\vec{A}\delta U ds + \\
&\int_S c_\infty^{-1}(\delta\vec{x}\cdot\hat{n})(\vec{d}\cdot\nabla p)ds + \int_\Omega \psi^T f d\Omega - \int_S \rho\hat{n}\cdot\delta\vec{v}(\psi_1 + \vec{v}\cdot\vec{\varphi})ds.
\end{aligned}
\tag{12}
$$

From Eq. (12) it follows that if the adjoint state obeys the following adjoint equations and boundary conditions

$$
\begin{aligned}
\vec{A}^T\cdot\nabla\psi &= 0 && \text{in } \Omega \\
\hat{n}\cdot\vec{\varphi} &= c_\infty^{-1}(\vec{d}\cdot\hat{n}) && \text{at } S \\
\psi^T\hat{n}\cdot\vec{A}\delta U &= 0 && \text{at } S_\infty
\end{aligned}
\tag{13}
$$

then the last two lines in Eq. (12) yield the (continuous) adjoint-based sensitivity derivatives

$$
\delta\int_S C_p\left(\hat{n}\cdot\vec{d}\right)ds = \int_\Omega \psi^T f d\Omega + \int_S c_\infty^{-1}(\delta\vec{x}\cdot\hat{n})(\vec{d}\cdot\nabla p)ds - \int_S w\rho(\psi_1 + \vec{v}\cdot\vec{\varphi})ds. \tag{14}
$$

Eq. (14) lacks the Leibniz term $\int_S (\delta\vec{x}\cdot\hat{n})\psi^T\nabla\cdot\vec{F}ds$, first introduced in (Kavvadias, et al., 2015) (see also (Lozano, 2017)), which is zero analytically. Finally, when far-field parameters (such as angle of attack) are considered as design variables, the contribution from the far-field integral $\int_{S_\infty}\psi^T\hat{n}\cdot\vec{A}\delta U ds$ should also be included in Eq. (14).

## 3. Using Green's functions to compute the analytic adjoint solution

The Green's function approach was completed with a closed-form solution for the adjoint quasi-1D Euler equations in (Giles & Pierce, 2001) and outlined for the 2D Euler equations in (Giles & Pierce, 1997), where four linearly independent source vectors representing mass, normal force, enthalpy and stagnation pressure perturbations, respectively, were considered. The first two reduce, for incompressible flow, to point mass sources and point vortices, respectively, which are well-known elementary solutions in potential flow; the third one yields zero lift or drag perturbation, and the final one was shown to lead to a non-uniform mass perturbation along the streamline downstream of the insertion point that would be responsible for an adjoint singularity along the incoming stagnation streamline.

Focusing now on 2D incompressible flow, suppose that we have three point perturbations $\delta U^{(j)}(\vec{x},\vec{x}_0)$, $j=1,...,3$, that obey the linearized equations

$$\nabla \cdot (\vec{A}\delta U^{(j)}(\vec{x},\vec{x}_0)) = f^{(j)}(\vec{x}_0)\delta(\vec{x}-\vec{x}_0), \tag{15}$$

with the following wall boundary conditions

$$\delta \vec{v}^{(j)}(\vec{x},\vec{x}_0)\cdot \hat{n} = 0, \;\; \vec{x}\in S \tag{16}$$

and far-field boundary conditions such that there is no perturbation to the inflow state. In (15), $\vec{x}$ is a generic point in the fluid domain, $\vec{x}_0$ is the location of the singularity and the gradient operator on the LHS of the equation acts on $\vec{x}$. If

$$I^{(j)}(\vec{x}_0) = c_\infty^{-1}\int_S \vec{d}\cdot \hat{n}\partial_U p\delta U^{(j)}(\vec{x},\vec{x}_0)ds \tag{17}$$

is the value of the linearized functional (lift or drag) corresponding to $\delta U^{(j)}(\vec{x},\vec{x}_0)$, then we can apply eqs. (12)-(14) to $\delta U^{(j)}(\vec{x},\vec{x}_0)$ and we get

$$I^{(j)}(\vec{x}_0) = \psi^T(\vec{x}_0) f^{(j)}(\vec{x}_0). \tag{18}$$

If the source vectors $f^{(j)}(\vec{x}_0)$ are linearly independent, Eq. (18) can be inverted to obtain the adjoint variables in terms of the linearized cost functions

$$\psi^T(\vec{x}_0) = \left(I^{(1)},I^{(2)},I^{(3)}\right)\cdot\left(f^{(1)} \mid f^{(2)} \mid f^{(3)}\right)^{-1}, \tag{19}$$

where $\left(f^{(1)} \mid f^{(2)} \mid f^{(3)}\right)$ is a matrix whose columns are the vectors $f^{(j)}$.

Following (Giles & Pierce, 1997), we now define 3 linearly independent perturbations, which we take to be a point source, a point vortex and a stagnation pressure perturbation at fixed pressure and local flow angle.

### 3.1. Point source

The first perturbation is a point mass source located at $\vec{x}_0$. In unbounded space, the response to this perturbation is given by the free-space solution

$$\delta U^{(1)}_{free} = \begin{pmatrix} \delta p^{(1)}_{free} \\ \rho\delta u^{(1)}_{free} \\ \rho\delta v^{(1)}_{free} \end{pmatrix} = \frac{\varepsilon}{2\pi}\begin{pmatrix} -(u\partial_x + v\partial_y)(\ln r) \\ \partial_x \ln r \\ \partial_y \ln r \end{pmatrix} \tag{20}$$

($r = |\vec{x}-\vec{x}_0|$ and $\varepsilon$ is an arbitrary constant point singularity strength that we introduce for convenience and that we will set to 1 in the end), which obeys the following linearized equations

$$\nabla\cdot\left(\vec{A}\delta U^{(1)}_{free}\right)=\begin{pmatrix}\rho\nabla\cdot\delta\vec{v}^{(1)}_{free}\\ \rho\vec{v}\cdot\nabla\delta u^{(1)}_{free}+\rho u\nabla\cdot\delta\vec{v}^{(1)}_{free}+\rho\delta\vec{v}^{(1)}_{free}\cdot\nabla u+\partial_x\delta p^{(1)}_{free}\\ \rho\vec{v}\cdot\nabla\delta v^{(1)}_{free}+\rho v\nabla\cdot\delta\vec{v}^{(1)}_{free}+\rho\delta\vec{v}^{(1)}_{free}\cdot\nabla v+\partial_y\delta p^{(1)}_{free}\end{pmatrix}= \tag{21}$$
$$\varepsilon\begin{pmatrix}1\\ u\\ v\end{pmatrix}\delta(\vec{x}-\vec{x}_0)+\frac{\varepsilon}{2\pi}\begin{pmatrix}0\\ (\partial_y u-\partial_x v)\partial_y\ln r\\ -(\partial_y u-\partial_x v)\partial_x\ln r\end{pmatrix},$$

where the second term on the right-hand side vanishes if the base flow is irrotational. Hence, the source vector is

$$f^{(1)}=\varepsilon\begin{pmatrix}1\\ u\\ v\end{pmatrix}. \tag{22}$$

Notice that $\psi^T f^{(1)}=\varepsilon(\psi_1+\vec{v}\cdot\vec{\varphi})$, so $I^{(1)}$ is related to the (continuous) adjoint-based lift/drag gradient (14).

The Green's function approach requires the value of the linearized functionals $I^{(1)}$, which are the lift and drag forces exerted by the point source on a body. In the presence of a body, the above functional form (20) only holds locally near the singularity, such that (22) still holds, but needs to be modified globally to account for the wall boundary condition. This is particularly simple in 2D potential flow using complex variable and the method of images, and allows us to compute the force on immersed bodies using Blasius theorem (Milne-Thomson, 1962). The simplest problem where the body is a circle is a standard academic exercise, and more general situations can be derived from this with suitable conformal transformations. The complex force on the cylinder turns out to be $w_I-q_\infty e^{-i\alpha}$, where $w_I$ is the induced velocity at the point singularity and $q_\infty e^{-i\alpha}$ is the free-stream complex velocity (here $q_\infty=\sqrt{{u_\infty}^2+{v_\infty}^2}$). An interesting fact is that the same expression yields the force on a body of arbitrary shape (Milne-Thomson, 1962), a result that is known as Lagally's theorem. Hence, the force exerted by the point source on the body is, keeping only terms linear in the source strength $\varepsilon$,

$$\delta D^{(1)}-i\delta L^{(1)}=\varepsilon e^{i\alpha}(u-iv-q_\infty e^{-i\alpha})+O(\varepsilon^2). \tag{23}$$

Here, $\delta D^{(1)}$ and $\delta L^{(1)}$ stand for the linear perturbations to drag and lift, respectively, due to the point source. Eq. (23) thus gives the linearized force exerted by a point source on a rigid body of arbitrary shape in terms of the undisturbed velocity at the source, $u-iv$. This is not the end of the story, though. Potential flows contain circulation, which can be freely adjusted *a priori*. For bodies with sharp trailing edges, the value of the circulation is fixed by the Kutta condition that asserts that the flow contains the precise amount of circulation to make the flow smooth at the trailing edge. The introduction of a point source disturbs the flow at the trailing edge, so the circulation has to be adjusted accordingly to preserve the smooth flow at the trailing edge. The extra circulation contributes an additional term to Eq. (23) that we shall calculate shortly. For blunt bodies, on the other hand, the circulation is not fixed by any smoothness requirement. However, it turns out that, in order to obtain consistent adjoint solutions, a Kutta condition has to be imposed as well, ensuring no perturbation to the position of the rear stagnation point (see section 3.2.1).

In order to obtain explicit formulae, we will consider an airfoil in the complex $z$ plane with a sharp trailing edge at $z = 1$. Let us suppose that the airfoil is transformed into a circle of radius $R$ and centered at $\zeta_0$ in the auxiliary complex $\zeta$ -plane by a conformal mapping $z = F(\zeta)$ (with $F(\zeta) \to 1$ as $|\zeta| \to \infty$), such that the point $\zeta_{te}$ on the circle is transformed into the trailing edge of the airfoil at $z = 1$. The (irrotational) base flow around the circle can be derived from the complex potential

$$\Phi(\zeta) = q_\infty e^{-i\alpha}(\zeta - \zeta_0) + q_\infty e^{i\alpha}\frac{R^2}{\zeta - \zeta_0} - \frac{i\Gamma_0}{2\pi}\ln(\zeta - \zeta_0). \tag{24}$$

The corresponding potential at a point $z$ on the airfoil plane is precisely $\Phi(\zeta(z))$. In the unperturbed flow, the circulation $\Gamma_0$ around the circle has to be fixed to the value

$$\Gamma_0 = -2\pi i q_\infty e^{-i\alpha}(\zeta_{te} - \zeta_0) + 2\pi i q_\infty e^{i\alpha}\frac{R^2}{\zeta_{te} - \zeta_0} \tag{25}$$

such that the flow around the circle has a stagnation point at $\zeta_{te}$, i.e.,

$$W(\zeta_{te}) = d\Phi / d\zeta\big|_{\zeta=\zeta_{te}} = q_\infty e^{-i\alpha} - q_\infty e^{i\alpha}\frac{R^2}{(\zeta_{te} - \zeta_0)^2} - \frac{i\Gamma_0}{2\pi}\frac{1}{\zeta_{te} - \zeta_0} = 0. \tag{26}$$

After inserting a point source at $\zeta_s$, the potential picks an extra contribution from the point source and its images with respect to the circle (Milne-Thomson, 1962)

$$\Phi_s(\zeta) = \frac{\varepsilon}{2\pi\rho}\ln\left(\zeta - \zeta_s\right) + \frac{\varepsilon}{2\pi\rho}\ln\left(\zeta - \zeta_0 - \frac{R^2}{\overline{\zeta_s} - \overline{\zeta_0}}\right) - \frac{\varepsilon}{2\pi\rho}\ln\left(\zeta - \zeta_0\right), \tag{27}$$

where $\overline{\zeta}$ denotes the complex conjugate of $\zeta$. $\Phi_s(\zeta)$ induces a non-zero velocity at $\zeta_{te}$,

$$W_s(\zeta_{te}) = \frac{\varepsilon}{2\pi\rho\left(\zeta_{te} - \zeta_s\right)} + \frac{\varepsilon}{2\pi\rho}\frac{1}{\zeta_{te} - \zeta_0 - \dfrac{R^2}{\overline{\zeta_s} - \overline{\zeta_0}}} - \frac{\varepsilon}{2\pi\rho\left(\zeta_{te} - \zeta_0\right)}, \tag{28}$$

which has to be cancelled with additional circulation

$$\delta\Gamma_0 = -2\pi i(\zeta_{te} - \zeta_0)W_s(\zeta_{te}). \tag{29}$$

This extra circulation produces an extra force that has to be added to (23)

$$
\begin{aligned}
&\delta D^{(1)} - i\delta L^{(1)} = i\rho q_\infty \delta\Gamma_0 + \varepsilon e^{i\alpha}(u - iv - q_\infty e^{-i\alpha}) + O(\varepsilon^2) = \\
&2\pi\rho q_\infty(\zeta_{te} - \zeta_0) W_s(\zeta_{te}) + \varepsilon e^{i\alpha}(u - iv - q_\infty e^{-i\alpha}) + O(\varepsilon^2) = \\
&-\varepsilon q_\infty \left( \frac{\zeta_{te} - \zeta_0}{\zeta_s - \zeta_{te}} - \frac{\overline{\zeta_{te}} - \overline{\zeta_0}}{\overline{\zeta_s} - \overline{\zeta_{te}}} \right) + \varepsilon e^{i\alpha}(u - iv - q_\infty e^{-i\alpha}) + O(\varepsilon^2)
\end{aligned}
\tag{30}
$$

Notice that the extra term is purely imaginary, so it only affects lift. Separating the real and imaginary parts, setting $\varepsilon = 1$ and dividing by $c_\infty$ to normalize, we get the following linearized functionals due to the point source

$$
\begin{aligned}
I_D^{(1)}(z(\zeta)) &= \frac{1}{c_\infty}\big(u(z(\zeta))\cos\alpha + v(z(\zeta))\sin\alpha - q_\infty\big), \\
I_L^{(1)}(z(\zeta)) &= \frac{1}{c_\infty}\big(v(z(\zeta))\cos\alpha - u(z(\zeta))\sin\alpha\big) - i\frac{q_\infty}{c_\infty}\left( \frac{\zeta_{te} - \zeta_0}{\zeta(z) - \zeta_{te}} - \frac{\overline{\zeta_{te}} - \overline{\zeta_0}}{\overline{\zeta}(z) - \overline{\zeta_{te}}} \right).
\end{aligned}
\tag{31}
$$

Notice that the extra contribution is independent of the initial circulation and blows up at the trailing edge, thus providing a precise mechanism (and an explanation) for the adjoint singularity at the trailing edge.

### 3.2. Point vortex

The second perturbation is a point vortex. In unbounded space it is simply

$$
\delta U_{free}^{(2)} = \begin{pmatrix} \delta p_{free}^{(2)} \\ \rho\delta u_{free}^{(2)} \\ \rho\delta v_{free}^{(2)} \end{pmatrix} = \frac{\varepsilon}{2\pi}\begin{pmatrix} (v\partial_x - u\partial_y)(\ln r) \\ \partial_y \ln r \\ -\partial_x \ln r \end{pmatrix}.
\tag{32}
$$

This solution obeys the linearized equation

$$
\begin{aligned}
\nabla\cdot\left(\vec{A}\delta U_{free}^{(2)}\right) &= \begin{pmatrix} \rho\nabla\cdot\delta\vec{v}_{free}^{(2)} \\ \rho\vec{v}\cdot\nabla\delta u_{free}^{(2)} + \rho u\nabla\cdot\delta\vec{v}_{free}^{(2)} + \rho\delta\vec{v}_{free}^{(2)}\cdot\nabla u + \partial_x\delta p_{free}^{(2)} \\ \rho\vec{v}\cdot\nabla\delta v_{free}^{(2)} + \rho v\nabla\cdot\delta\vec{v}_{free}^{(2)} + \rho\delta\vec{v}_{free}^{(2)}\cdot\nabla v + \partial_y\delta p_{free}^{(2)} \end{pmatrix} = \\
&\varepsilon\begin{pmatrix} 0 \\ v \\ -u \end{pmatrix}\delta(\vec{x} - \vec{x}_0) + \frac{\varepsilon}{2\pi}\begin{pmatrix} 0 \\ (\partial_x v - \partial_y u)\partial_x \ln r \\ (\partial_x v - \partial_y u)\partial_y \ln r \end{pmatrix},
\end{aligned}
\tag{33}
$$

where the extra term vanishes if the base flow is irrotational. Hence, the source vector is

$$
f^{(2)} = \varepsilon\begin{pmatrix} 0 \\ v \\ -u \end{pmatrix}.
\tag{34}
$$

Notice that $\psi^T f^{(2)} = \varepsilon(v\psi_2 - u\psi_3)$, so $I^{(2)}$ approaches $q\vec{n}_S \cdot \vec{\varphi}$ as the vortex approaches the wall. Its value is thus constrained by the adjoint wall b.c. $\hat{n}_S \cdot \vec{\varphi} = c_\infty{}^{-1}(\vec{d} \cdot \hat{n}_S)$ (Giles & Pierce, 1997).

Once again, in the presence of a body the above functional form needs to be modified to account for the wall boundary condition. The correct expression can then be used to compute the force exerted by the vortex on a body, which poses no significant difficulties relative to the source case, so we just quote the results. The linearized complex force, taking into account the Kutta condition, is

$$\begin{aligned}\delta D^{(2)} - i\delta L^{(2)} &= 2\pi\rho q_\infty(\zeta_{te} - \zeta_0)W_v(\zeta_{te}) + i\varepsilon e^{i\alpha}(u - iv - q_\infty e^{-i\alpha}) + O(\varepsilon^2) = \\ &-i\varepsilon q_\infty\left(\frac{\zeta_{te} - \zeta_0}{\zeta_s - \zeta_{te}} + \frac{\overline{\zeta_{te}} - \overline{\zeta_0}}{\overline{\overline{\zeta_s}}  - \overline{\overline{\zeta_{te}}}}\right) + i\varepsilon e^{i\alpha}(u - iv - q_\infty e^{-i\alpha}) + O(\varepsilon^2),\end{aligned} \tag{35}$$

where $W_v(\zeta_{te})$ is the complex velocity induced by the vortex (and its images) at $\zeta_{te}$. Hence, the properly normalized linearized functionals due to the vortex are (setting $\varepsilon = 1$ )

$$\begin{aligned}I_D^{(2)}(\zeta) &= \frac{1}{c_\infty}\big(v(z(\zeta))\cos\alpha - u(z(\zeta))\sin\alpha\big) \\ I_L^{(2)}(z(\zeta)) &= -\frac{1}{c_\infty}\big(u(z(\zeta))\cos\alpha + v(z(\zeta))\sin\alpha - q_\infty\big) + \frac{q_\infty}{c_\infty}\left(\frac{\zeta_{te} - \zeta_0}{\zeta_s - \zeta_{te}} + \frac{\overline{\zeta_{te}} - \overline{\zeta_0}}{\overline{\overline{\zeta_s}} - \overline{\overline{\zeta_{te}}}}\right).\end{aligned} \tag{36}$$

Notice that the extra term diverges as the vortex approaches the wall. However, along the wall we have $|\zeta_s - \zeta_0| = R$ and, thus,

$$\frac{\zeta_{te} - \zeta_0}{\zeta_s - \zeta_{te}} + \frac{\overline{\zeta_{te}} - \overline{\zeta_0}}{\overline{\overline{\zeta_s}} - \overline{\overline{\zeta_{te}}}} = -1, \tag{37}$$

so that

$$I_L^{(2)}\Big|_{wall} = -\frac{1}{c_\infty}\big(u\cos\alpha + v\sin\alpha\big) \tag{38}$$

as required by the adjoint wall b.c.

### 3.2.1. *Blunt Bodies*

For blunt bodies (circle, ellipse) there is no need to impose a Kutta condition, so the circulation is *a priori* arbitrary and the above extra terms disappear. However, the resulting forces are not consistent with the adjoint approach (as we have discussed above, the force exerted by a point vortex has to approach the wall in a certain fashion so that the adjoint wall b.c. is obeyed) and the Green's function approach leads to the wrong lift-based adjoint solution. It turns out that to obtain a sensible solution it is necessary to assume that an adjoint-consistent linearization requires a Kutta condition such that the rear stagnation point is not disturbed by the point singularities. This requirement results in the extra forces described above, and the linearized functionals are precisely (31) and (36) and lead to an analytic adjoint solution that obeys the adjoint equations and boundary conditions and that agrees with numerical computations (see for example the adjoint solution for a 6:1 ellipse presented in (Lozano & Ponsin, 2021)).

### 3.3. Change in stagnation pressure at fixed $p$

For the final perturbation we choose, in analogy with (Giles & Pierce, 1997), a perturbation to the stagnation pressure $p_0 = p + \frac{1}{2}\rho q^2$ (where $q^2 = u^2 + v^2$) at fixed values of the pressure and the local flow angle $\phi$. As stagnation pressure is convected along streamlines, we expect the solution to be concentrated along the streamline downstream of the insertion point $\vec{x}_0$. Following (Giles & Pierce, 1997), we use curvilinear streamline coordinates in which $s$ is the distance along the streamline downstream of the insertion point and $n$ is the coordinate perpendicular to the streamline. In this coordinates, the perturbation is, in a first approximation,

$$\varepsilon H(s)\delta(n)\frac{1}{q}\left(\frac{\partial U}{\partial p_0}\right)_{p,\phi}(\vec{x}) = \frac{\varepsilon}{q^3}\begin{pmatrix} 0 \\ u \\ v \end{pmatrix}(\vec{x})H(s)\delta(n), \tag{39}$$

where $H(s)$ is the Heaviside step function and $\delta(n)$ is the Dirac delta function. However, as explained in (Giles & Pierce, 1997), this is not the full form of the solution. This fact can be understood most simply by noting that (39) does not obey the linearized flow equations. In fact,

$$\nabla\cdot\left[\frac{\varepsilon}{q}\vec{A}\left(\frac{\partial U}{\partial p_0}\right)_{p,\phi} H(s)\delta(n)\right] = \nabla\cdot\left[\hat{s}f^{(3)}H(s)\delta(n)\right], \tag{40}$$

where $\hat{s}$ is the unit vector in the flow direction and

$$f^{(3)} = \frac{\varepsilon}{q}\left(\frac{\partial \vec{F}\cdot\hat{s}}{\partial p_0}\right)_{p,\phi} = \frac{\varepsilon}{q^2}\begin{pmatrix} 1 \\ 2u \\ 2v \end{pmatrix}. \tag{41}$$

To proceed further, we need to express $\nabla$ in streamline coordinates. While for smooth functions $\nabla\cdot[\hat{s}w] = \partial_s w + w\partial_n\phi$, when delta functions are present one needs to proceed more carefully. To this end, we follow (Emanuel, 2000) and define coordinates $\xi_1$ along the streamline and $\xi_2$ that parameterizes the lines orthogonal to $\xi_1$. The $\xi_i$ coordinates are related to the streamline coordinates as $ds = h_1 d\xi_1$ and $dn = h_2 d\xi_2$, where $h_i = \|\partial\vec{x}/\partial\xi_i\|$ are the scale factors. In terms of $(\xi_1, \xi_2)$, the divergence operator acting on a vector $\vec{w}$ is

$$\nabla\cdot\vec{w} = \frac{1}{h_1 h_2}\frac{\partial}{\partial\xi_1}\left(h_2\vec{w}\cdot\hat{e}_1\right) + \frac{1}{h_1 h_2}\frac{\partial}{\partial\xi_2}\left(h_1\vec{w}\cdot\hat{e}_2\right), \tag{42}$$

where $\hat{e}_i = h_i^{-1}\partial\vec{x} / \partial\xi_i$ are the local curvilinear unit basis vectors (note that $\hat{e}_1 \equiv \hat{s}$ in earlier notation). Finally, we need to consider the Heaviside and Dirac delta functions. Since $\iint dxdy = \iint h_1 h_2 d\xi_1 d\xi_2$ , we have

$$\delta(\vec{x}) \equiv \delta(x)\delta(y) = \delta(\xi_1)\delta(\xi_2) / h_1 h_2 \tag{43}$$

and

$$H(s)\delta(n) \equiv H(\xi_1)\delta(\xi_2) / h_2. \tag{44}$$

Hence,

$$\begin{aligned}
&\nabla \cdot ( f^{(3)} \hat{s} H(s)\delta(n)) = \frac{1}{h_1 h_2} \frac{\partial}{\partial\xi_1}\left(H(\xi_1)\delta(\xi_2) f^{(3)}\right) = \\
&\frac{1}{h_1 h_2}\delta(\xi_1)\delta(\xi_2) f^{(3)} + \frac{H(\xi_1)\delta(\xi_2)}{h_2}\frac{1}{h_1}\frac{\partial}{\partial\xi_1} f^{(3)} = \\
&\delta(\vec{x}-\vec{x}_0) f^{(3)} + H(s)\delta(n)\frac{\partial}{\partial s} f^{(3)}.
\end{aligned} \tag{45}$$

Using $(u,v) = (q\cos\phi, q\sin\phi)$, the last term on the right-hand side can be further expanded as

$$\frac{\partial}{\partial s} f^{(3)} = f^{(1)}\partial_s q^{-2} - \frac{2}{q^2} f^{(2)}\partial_s \phi, \tag{46}$$

where

$$f^{(1)} = \varepsilon \begin{pmatrix} 1 \\ u \\ v \end{pmatrix}, \qquad f^{(2)} = \varepsilon \begin{pmatrix} 0 \\ v \\ -u \end{pmatrix} \tag{47}$$

are the source vectors for the point source and vortex, respectively. Hence, we can write the full solution as

$$\begin{aligned}
&\delta U^{(3)}(\vec{x},\vec{x}_0) = \varepsilon H(s)\delta(n)\frac{1}{q}\left(\frac{\partial U}{\partial p_0}\right)_{p,\phi}(\vec{x}) \\
&-\iint h_1 h_2 d\xi_1' d\xi_2' \frac{H(\xi_1'-\xi_1^0)\delta(\xi_2'-\xi_2^0)}{h_2}\partial_s q^{-2}(\vec{x}')\delta U^{(1)}(\vec{x},\vec{x}') \\
&+\iint h_1 h_2 d\xi_1' d\xi_2' \frac{H(\xi_1'-\xi_1^0)\delta(\xi_2'-\xi_2^0)}{h_2}\frac{2}{q^2}\partial_s \phi(\vec{x}')\delta U^{(2)}(\vec{x},\vec{x}') = \\
&\varepsilon H(s)\delta(n)\frac{1}{q}\left(\frac{\partial U}{\partial p_0}\right)_{p,\phi}(\vec{x}) - \int_0^\infty ds'\partial_s q^{-2}(\vec{x}(s'))\delta U^{(1)}(\vec{x},\vec{x}(s')) \\
&+\int_0^\infty ds'\frac{2}{q^2}\partial_s\phi(\vec{x}(s'))\delta U^{(2)}(\vec{x},\vec{x}(s')),
\end{aligned} \tag{48}$$

where $\delta U^{(1)}$, $\delta U^{(2)}$ are the complete point source and point vortex Green's functions (i.e., respecting the wall boundary conditions), which reduce to (20) and (32), respectively, in the vicinity of the singularity. We can check that Eq. (48) obeys

$$\nabla \cdot (\vec{A}\delta U^{(3)}(\vec{x},\vec{x}_0)) = f^{(3)}(\vec{x})\delta(\vec{x}-\vec{x}_0) = f^{(3)}(\vec{x}_0)\delta(\vec{x}-\vec{x}_0), \tag{49}$$

as expected. Notice that $\left(\dfrac{\partial \rho q}{\partial p_0}\right)_{p,\phi} = \dfrac{1}{q}$ so that $\partial_s q^{-2} = \partial_s \left[\dfrac{1}{q}\left(\dfrac{\partial \rho q}{\partial p_0}\right)_{p,\phi}\right] = \rho \partial_s \tilde{m}$ where $\tilde{m}$ is the linearized mass flux perturbation defined in (Giles & Pierce, 1997), and thus the first integral on the right-hand side of (48) is actually the integral derived in (Giles & Pierce, 1997). The second integral also has an analog in the compressible case, so the solution shown in (Giles & Pierce, 1997) is apparently incomplete.

For pressure-based functionals, the first term in (48) does not contribute to the linearized functional, which can then be written as

$$I^{(3)}(\vec{x}_0) = -\int_0^\infty ds \partial_s q^{-2}(\vec{x}(s)) I^{(1)}(\vec{x}(s)) + \int_0^\infty ds \frac{2}{q^2} \partial_s \phi(\vec{x}(s)) I^{(2)}(\vec{x}(s)) \tag{50}$$

Again, the first integral on the RHS of (50) is equivalent to the corresponding integral in (Giles & Pierce, 1997), where it was used to argue that there is a potential divergence along the incoming stagnation streamline. Whether this singularity is actually present or not depends on the cost function and the flow conditions. Numerical tests show that lift-adjoint solutions seem to have these singularities even in incompressible and irrotational flow, while drag-adjoint solutions only show them in certain transonic cases (Lozano & Ponsin, 2021).

We can now substitute (31) and (36) into (50) to compute the linearized functional corresponding to the stagnation pressure point perturbation (48). For drag, we get

$$I_D^{(3)} = -\frac{1}{q^2 q_\infty c_\infty}(\vec{v}-\vec{v}_\infty)^2, \tag{51}$$

while for lift the result is

$$I_L^{(3)}(\vec{x}_0) = -\frac{2}{q^2 c_\infty}(u \sin\alpha - v\cos\alpha) + \frac{q_\infty}{c_\infty}\Xi \tag{52}$$

where

$$\Xi = -\int_0^\infty ds \partial_s q^{-2}(\vec{x}(s)) \Upsilon^{(1)}(\vec{x}(s)) + 2\int_0^\infty ds \frac{1}{q^2} \partial_s \phi(\vec{x}(s))(1+\Upsilon^{(2)}(\vec{x}(s))), \tag{53}$$

and

$$\Upsilon^{(1)}(z) = -i\left( \frac{\zeta_{te} - \zeta_0}{\zeta(z) - \zeta_{te}} - \frac{\overline{\zeta_{te}} - \overline{\zeta_0}}{\overline{\zeta}(z) - \overline{\zeta_{te}}} \right) = 2\frac{X_{te}\left(Y_0 - Y\right) + X_0\left(Y - Y_{te}\right) + X\left(Y_{te} - Y_0\right)}{\left(X - X_{te}\right)^2 + \left(Y - Y_{te}\right)^2},$$

$$\Upsilon^{(2)}(z) = \frac{\zeta_{te} - \zeta_0}{\zeta(z) - \zeta_{te}} + \frac{\overline{\zeta_{te}} - \overline{\zeta_0}}{\overline{\zeta}(z) - \overline{\zeta_{te}}} = \tag{54}$$

$$2\frac{\left(X - X_0\right)\left(X - X_{te}\right) + \left(Y - Y_0\right)\left(Y - Y_{te}\right) - \left(X - X_{te}\right)^2 - \left(Y - Y_{te}\right)^2}{\left(X - X_{te}\right)^2 + \left(Y - Y_{te}\right)^2}.$$

In (53), the integration is carried out along the streamline downstream of $\vec{x}_0$. Finally, $X$ and $Y$ in (54) are the Cartesian coordinates on the circle plane, in terms of which $\zeta = X + iY$.

## 4. Analytic adjoint solutions

Using (19), (22), (31), (34), (36), (41), (51) and (52), we can compute the analytic adjoint solutions as

$$(\psi_1, \psi_2, \psi_3) = (I^{(1)}, I^{(2)}, I^{(3)}) \begin{pmatrix} 1 & 0 & q^{-2} \\ u & v & 2q^{-2}u \\ v & -u & 2q^{-2}v \end{pmatrix}^{-1} =$$

$$(I^{(1)}, I^{(2)}, I^{(3)}) \begin{pmatrix} 2 & -q^{-2}u & -q^{-2}v \\ 0 & q^{-2}v & -q^{-2}u \\ -q^2 & u & v \end{pmatrix}. \tag{55}$$

For **drag** we get

$$\begin{pmatrix} \psi_1 \\ \psi_2 \\ \psi_3 \end{pmatrix}_{Drag} = \frac{1}{c_\infty q_\infty} \begin{pmatrix} q^2 - q_\infty{}^2 \\ q_\infty \cos\alpha - u \\ q_\infty \sin\alpha - v \end{pmatrix} \tag{56}$$

which is exactly the analytic drag adjoint solution found in (Lozano & Ponsin, 2021). It can be checked that it obeys the adjoint equation and boundary conditions. Notice that no assumption has been made about the lift of the base flow solution. Hence, this two-dimensional solution should be valid for lifting and non-lifting flows alike, contrary to what was claimed in (Lozano & Ponsin, 2021), where the validity of the solution was restricted to non-lifting flows based on far-field behavior. In 2D, incompressible flow past a lifting airfoil is still irrotational. The only difference between a lifting and a non-lifting flow is in the order of the subleading terms as $r \to \infty$

$$\begin{aligned} \vec{v} &= \vec{v}_\infty + O(1/r^2) \quad (non-lifting), \\ \vec{v} &= \vec{v}_\infty + O(1/r) \quad (lifting). \end{aligned} \tag{57}$$

Adjoint b.c. at the outer boundary demand

$$\int_{S_\infty} \psi^T \hat{n}\cdot\vec{A}\delta U ds = 0 \tag{58}$$

for perturbations $\delta U$ respecting the far-field state. For incompressible flow, we have

$$\psi_{drag}^T \hat{n}\cdot\vec{A}\delta U = \frac{\rho q_\infty \Gamma_0}{2\pi r}(\delta u \sin\alpha - \delta v\cos\alpha) + O(1/r^2), \tag{59}$$

where $\Gamma_0$ is the circulation of the base flow, so if $\delta u, \delta v$ tend to zero at the far-field (which is a reasonable assumption as otherwise the free-stream state would be perturbed), then the above term is at least $O(1/r^2)$ and the far-field integral vanishes, thus guaranteeing that $\psi_{drag}$ is also a valid solution for lifting flows. The discrepancies observed in (Lozano & Ponsin, 2021) can then be attributed to the dependence of the numerical solution on the distance from the body to the far-field (20 chord lengths for the meshes used in (Lozano & Ponsin, 2021)).

On the other hand, for **lift** we get the result

$$\begin{pmatrix}\psi_1\\ \psi_2\\ \psi_3\end{pmatrix}_{Lift} = \frac{q_\infty}{c_\infty}\begin{pmatrix} 2\Upsilon^{(1)} - q^2\Xi \\ -\frac{\sin\alpha}{q_\infty} + u\Xi - \frac{u}{q^2}\Upsilon^{(1)} + \frac{v}{q^2}\left(1+\Upsilon^{(2)}\right) \\ \frac{\cos\alpha}{q_\infty} + v\Xi - \frac{v}{q^2}\Upsilon^{(1)} - \frac{u}{q^2}\left(1+\Upsilon^{(2)}\right)\end{pmatrix}, \tag{60}$$

It can be checked (with a long derivation that can be conveniently alleviated with a symbolic manipulation software) that Eq. (60) obeys the adjoint equations, for which one needs to use the following identities

$$\begin{aligned}\partial_x \Upsilon^{(2)}(z) &= \partial_y \Upsilon^{(1)}(z),\\ \partial_y \Upsilon^{(2)}(z) &= -\partial_x \Upsilon^{(1)}(z),\end{aligned} \tag{61}$$

which follow from the holomorphicity of $\Upsilon^{(2)}(z) + i\Upsilon^{(1)}(z)$, as well as

$$\vec{v}\cdot\nabla\Xi = q\partial_s q^{-2}\Upsilon^{(1)} - q\frac{2}{q^2}\partial_s\phi(1+\Upsilon^{(2)}) = -\frac{2}{q^3}\Upsilon^{(1)}\vec{v}\cdot\nabla q - \frac{2}{q^2}\vec{v}\cdot\nabla\phi(1+\Upsilon^{(2)}) \tag{62}$$

and the incompressibility and irrotationality of the base flow. Likewise, (60) also obeys the wall boundary conditions,

$$\begin{aligned}&c_\infty(\vec{\varphi}\cdot\hat{n}) = c_\infty(n_x\psi_2 + n_y\psi_3) =\\ &(-\sin\alpha,\cos\alpha)\cdot\hat{n} + q_\infty(\vec{v}\cdot\hat{n})(\Xi - q^{-2}\Upsilon^{(1)}) + \frac{q_\infty}{q^2}\left(1+\Upsilon^{(2)}\right)(v,-u)\cdot\hat{n} =\\ &(-\sin\alpha,\cos\alpha)\cdot\hat{n},\end{aligned} \tag{63}$$

since $\vec{v}\cdot\hat{n}=0$ and $1+\Upsilon^{(2)}=0$ at the wall.

### 4.1. Behavior of $\Xi$

The streamline integral eq. (53) represents the contribution to lift, due to the Kutta condition, of a unit point perturbation to the stagnation pressure. This perturbation generates secondary point sources and vortices along the streamline downstream of the insertion point. The contributions of these secondary perturbations to the lift (see (31), (36)) diverge near the trailing edge, which gives rise to a singular behavior of $\Xi$ near the dividing streamline upstream of the trailing edge or rear stagnation point (which includes the wall and the incoming stagnation streamline). Along the stagnation streamline upstream of the body, the integral is not defined, as the streamline splits up in two at the stagnation point, while along the wall the integral diverges. For a circle, for example, the integral along the wall is

$$\Xi(\vec{x}_0)\sim\int_0^{\theta_0} d\theta\cos^2(\theta/2)\cot\theta\csc^3\theta, \tag{64}$$

where $\vec{x}_0=(R\cos\theta_0, R\sin\theta_0)$, which diverges as $\theta^{-3}$ at the rear stagnation point ($\theta=0$). Away from the dividing streamline, the integral grows unbounded as the corresponding streamline approaches the dividing streamline. In order to investigate the character of the divergence, we consider a simplified setup involving flow past a wedge of half-angle $\tau/2$ (Figure 1).

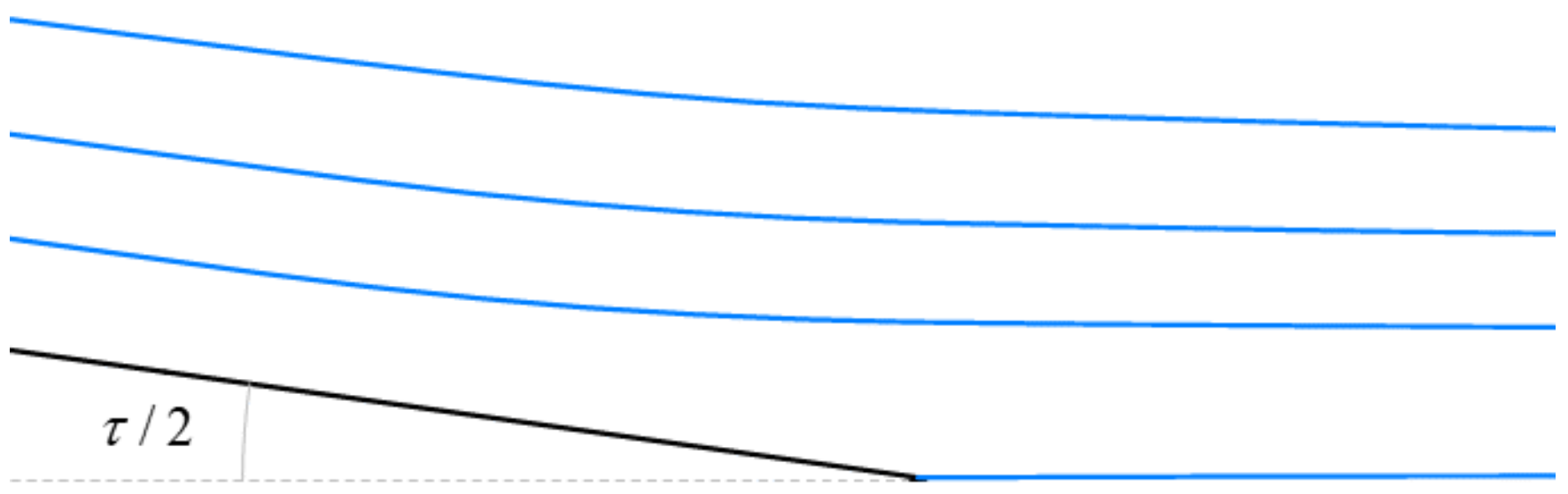

**Figure 1.** Streamlines for potential flow past a wedge of angle $\tau$.

The asymptotic form of the complex potential for this case is $z^{2/k}$, where $k=2-\tau/\pi$, so that stream function reads, in polar coordinates, $\psi=r^{2/k}\sin(2\theta/k)$, from which it follows that the minimum distance from a particular streamline to the trailing edge is $r_{\min}\sim\psi^{k/2}$, while the distance from a given point on the streamline to the dividing streamline is $d\sim\psi\sim r_{\min}{}^{2/k}$. Likewise, the flow speed behaves as $q=O(r^{\frac{2}{k}-1})$, so $\partial_s q=O(r^{\frac{2}{k}-2})$, and likewise $\partial_s\phi=O(r^{-1})$. Finally, $\zeta\sim z^{1/k}$ and, thus, $\Upsilon^{(1)}=O(r^{-\frac{1}{k}})$ and $\Upsilon^{(2)}=O(r^{-\frac{1}{k}})$, so integrating along a streamline,

$$\Xi\sim r_{\min}^{2-\frac{5}{k}}\sim r_{\min}^{-\frac{2\tau/\pi+1}{2-\tau/\pi}}, \tag{65}$$

which depends on the trailing edge wedge angle $\tau$. Notice that the exponent is always negative for $0 < \tau \leq \pi$. It is thus clear that $\Xi$ has a potential divergence as the streamline approaches the dividing streamline. Let us consider three different scenarios in terms of the relative position of $\vec{x}_0$.

1. $\vec{x}_0$ approaching the rear singularity (trailing edge or rear stagnation point). As $\vec{x}_0$ approaches the rear singularity (Figure 2), the distance to the rear singularity behaves as $r \sim r_{\min}$ asymptotically. Hence, $\Xi(\vec{x}_0)$ diverges as $r^{2-\frac{5}{k}} \sim r^{-\frac{2\tau/\pi+1}{2-\tau/\pi}}$ as $\vec{x}_0$ approaches the rear singularity. This singular behavior had been already anticipated (including the dependence on the trailing edge angle) in (Giles & Pierce, 1997), but no concrete predictions were made for the exponent. A $1/r_{\min}$ singularity at the rear stagnation point was predicted in (Lozano & Ponsin, 2021) for blunt bodies in potential flow using the form of the streamline integral derived in (Giles & Pierce, 1997), while the actual value for this case (for which $\tau = \pi \Rightarrow k = 1$) is $\Xi \sim r^{-3}$.

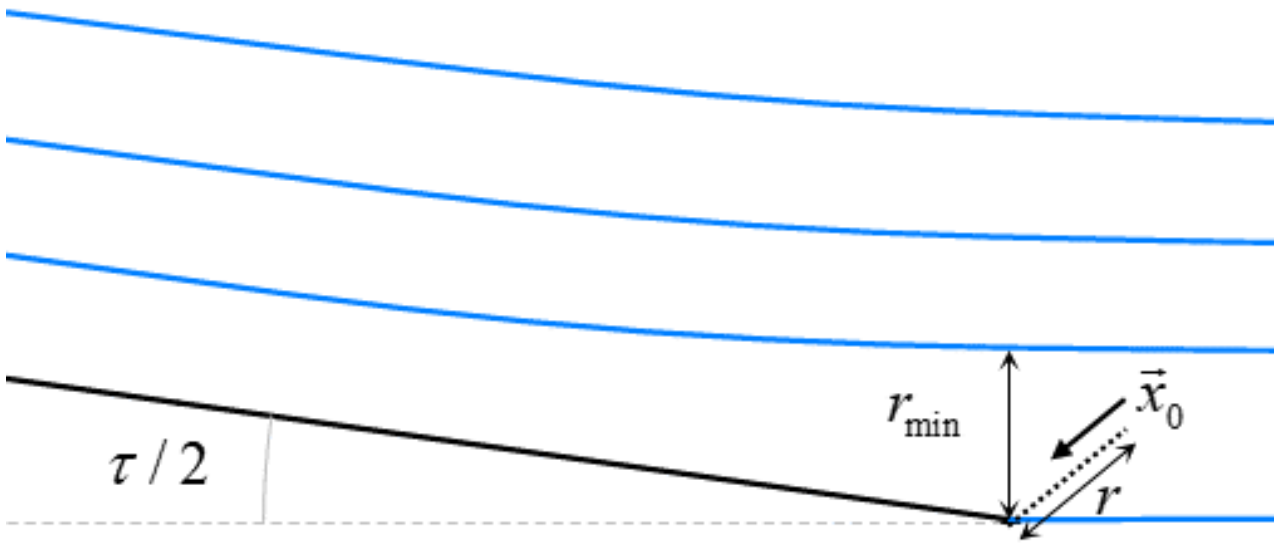

**Figure 2.** Set-up for case 1.

2. $\vec{x}_0$ upstream of the rear singularity (Figure 3). As $\vec{x}_0$ approaches the dividing streamline upstream of the rear singularity, $\Xi(\vec{x}_0)$ diverges as $r_{\min}^{2-\frac{5}{k}} \sim d^{k-5/2} \sim 1/d^{\tau/\pi+1/2}$, where $d$ is the distance from $\vec{x}_0$ to the dividing streamline (incoming stagnation streamline or wall). Notice that the singularity exponent is not universal since it depends on the trailing edge angle $\tau$, and reduces to the inverse square-root behavior predicted in (Giles & Pierce, 1997) for cusped trailing edges (for which $\tau = 0$).

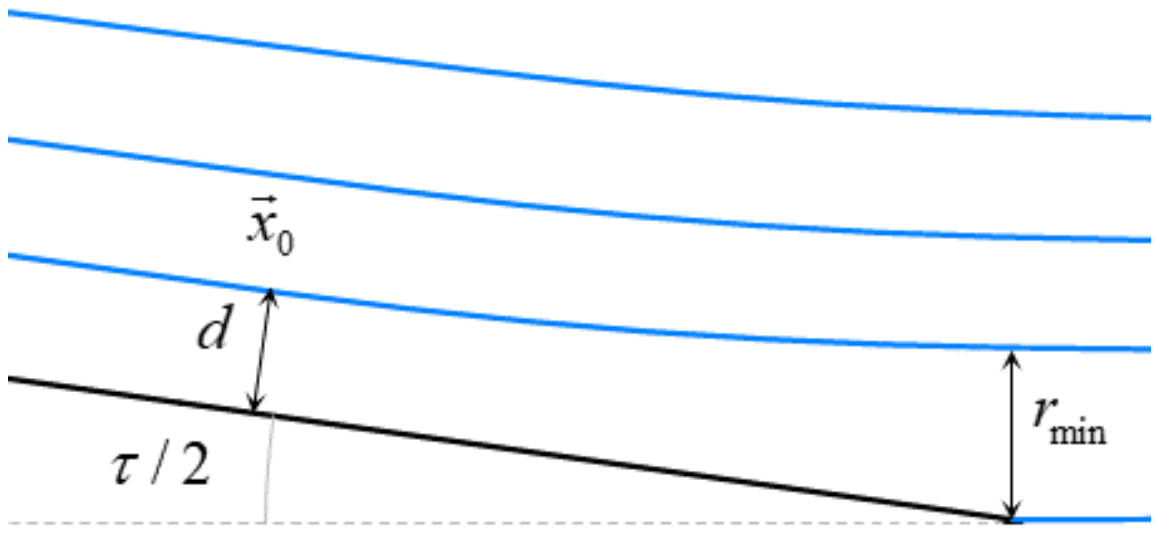

**Figure 3.** Set-up for case 2.

3. As $\vec{x}_0$ approaches the rear stagnation streamline downstream of the rear singularity (Figure 4), $\Xi(\vec{x}_0)$ does not diverge (the minimum distance from the streamline to the trailing edge is now $|\vec{x}_0 - \vec{x}_{te}|$, which is always greater than zero) but rather behaves as $\Xi(\vec{x}_0) \sim d$, where $d$ is the distance from $\vec{x}_0$ to the dividing streamline.

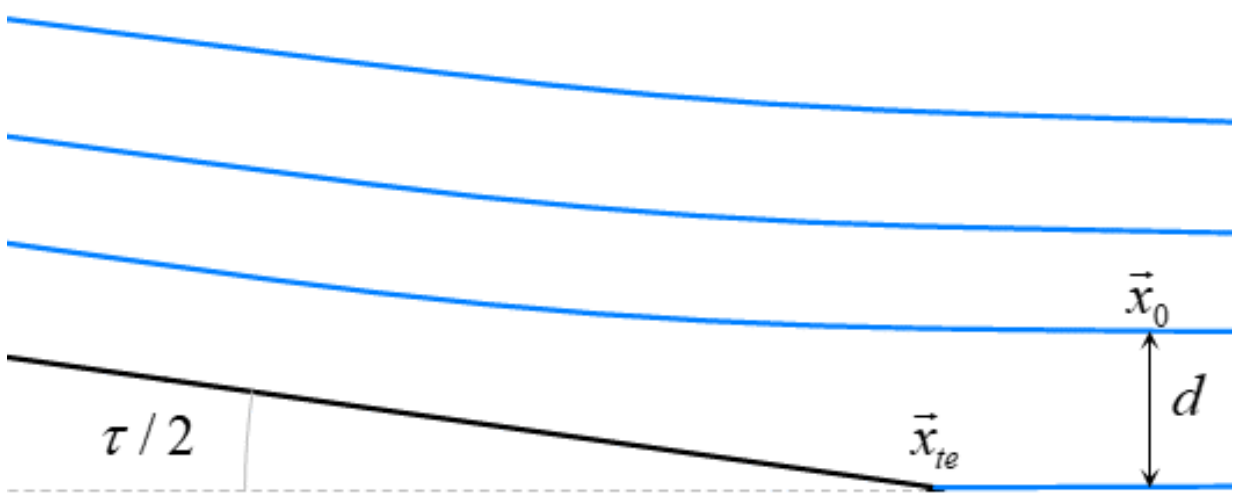

**Figure 4.** Set-up for case 3.

The results of this asymptotic analysis are confirmed by actual computations (see section 5). This suggests that the origin of the divergence along the dividing streamline is the singularity at the trailing edge/rear stagnation point. This can be put on a firmer basis by noting that, as it will be shown later on, while the integral does get a small contribution from the leading edge region, it is very approximately constant upstream of the rear singularity and vanishingly small downstream. Likewise, the integral behaves in exactly the same fashion at both the incoming stagnation streamline and the wall, with an exponent that depends on the local geometry around the rear singularity.

#### 4.1.1. *Influence of $\Xi$ on the sensitivity derivatives and the adjoint boundary conditions*

The divergences of $\Xi$ (and $\Upsilon^{(1)}$ and $\Upsilon^{(2)}$) are propagated to the lift-based adjoint variables (60). However, two linear combinations of the adjoint variables,

$$\begin{aligned} I^{(1)} &= \psi_1 + \vec{v}\cdot(\psi_2,\psi_3) = (q_\infty \Upsilon^{(1)} - u\sin\alpha + v\cos\alpha)/c_\infty, \\ I^{(2)} &= v\psi_2 - u\psi_3 = (-v\sin\alpha - u\cos\alpha + q_\infty(1+\Upsilon^{(2)}))/c_\infty, \end{aligned} \tag{66}$$

which yield the linearized perturbations to the lift due to a point source and a point vortex, respectively (Giles & Pierce, 1997), remain finite except at the trailing edge. This fact is relevant since the adjoint variables enter the (continuous) adjoint-based sensitivity derivatives (14) precisely through $I^{(1)}$, which protects the sensitivities from diverging with mesh refinement (a fact that is well established numerically, see e.g. (Lozano, 2019) (Lozano, 2017) (Lozano, 2012) (Lozano & Ponsin, 2022)). Likewise, $I^{(2)}$ approaches $q\hat{n}\cdot(\psi_2,\psi_3)$ as the point approaches the wall and is thus directly related to the adjoint boundary condition, which is thus respected by the analytical solution and protected upon mesh refinement (Lozano, 2019). Finally, the divergence along the dividing streamline is encapsulated into a third (linearly independent) combination

$$I^{(3)} = (\psi_1 + 2u\psi_2 + 2v\psi_3)/q^2 = (q_\infty \Xi - 2q^{-2}u\sin\alpha + 2q^{-2}v\cos\alpha)/c_\infty, \tag{67}$$

which yields the linearized perturbation to the lift due to a unit stagnation pressure point perturbation.

It is important to point out, however, that the above split of the adjoint solution space into protected and unprotected terms (as regards the dividing streamline singularity) is arbitrary and based on our choice of point perturbations (inherited from (Giles & Pierce, 1997)). In fact, it turns out that any linear combination

$$a\psi_1 + b\psi_2 + c\psi_3, \tag{68}$$

which corresponds to a point perturbation with source vector $f = (a,b,c)^T$, obeying $-aq^2 + bu + cv = 0$, does not contain the streamline integral $\Xi$ and is thus protected from the dividing streamline singularity. Conversely, if $-aq^2 + bu + cv \neq 0$ then the corresponding perturbation is singular along the dividing streamline. One such example is $f = (0,u,v)^T$, which is related to the drag Eulerlet (Chadwick, et al., 2019), a Green's function of the incompressible Euler equations linearized around a uniform flow which carries one unit of drag.

## 5. Sample solutions

In order to illustrate the results presented above, we present now the results of direct evaluation of the adjoint analytical solutions in two cases: blunt body (a circle) and an airfoil with a sharp non-cusped trailing edge.

### 5.1. Circle

As explained above, obtaining meaningful adjoint solutions for bodies without sharp trailing edges such as the circle requires a Kutta condition to guarantee that the perturbations preserve the location of the rear stagnation point. As a consequence of this fact, the solutions (56) and (60) are applicable here, resulting in the analytic drag-based adjoint solution shown in Figure 5

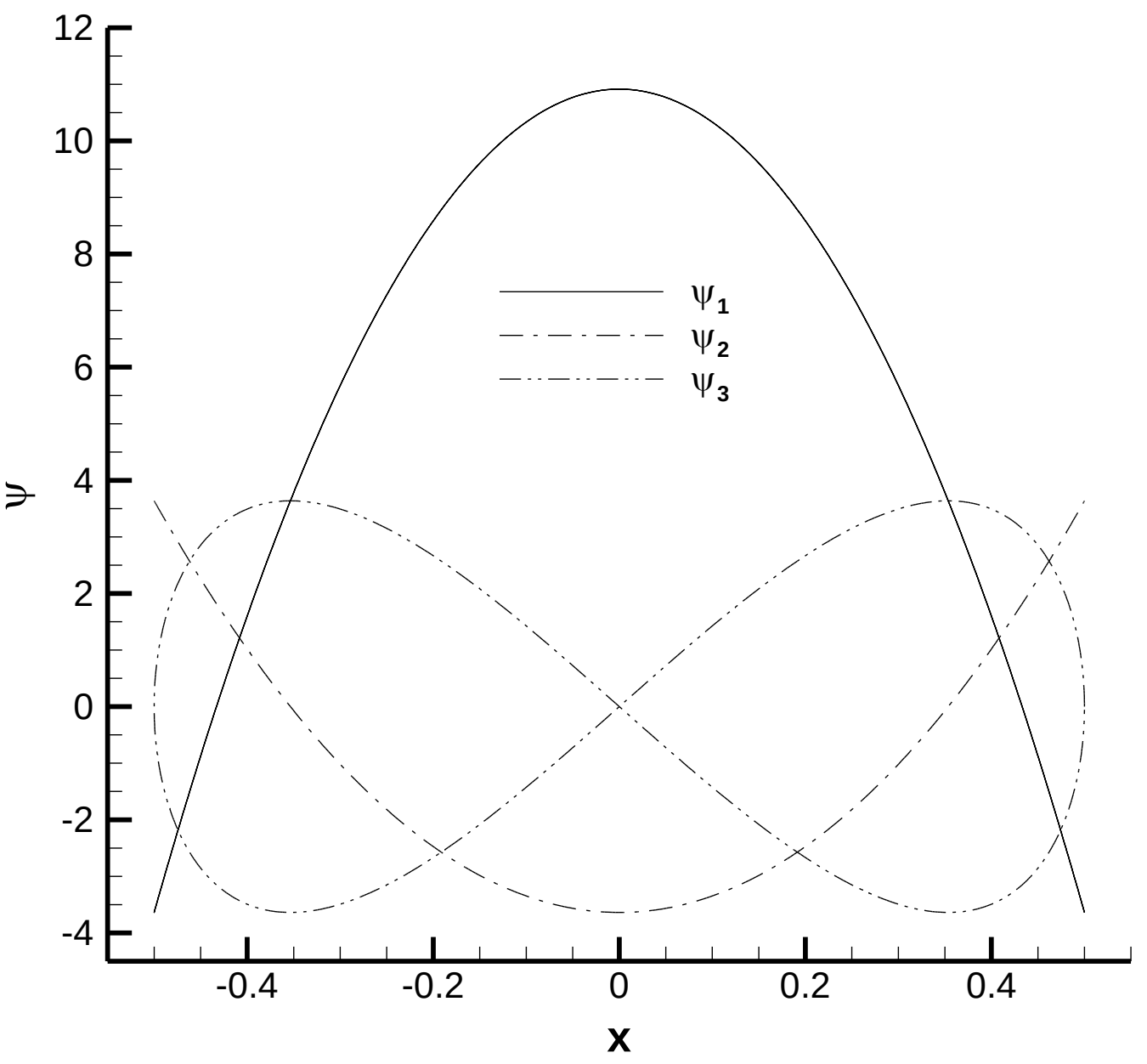

**Figure 5.** Analytic drag-based adjoint solution at the wall for incompressible, inviscid flow past a circle with $\Gamma_0 = 0$.

Here and in what follows, the normalization of the adjoint variables is given by (56) and (60). For the flow variables, dimensions are fixed such that $\rho_\infty = 1 = q_\infty$, while spatial coordinates are normalized with the reference length $\ell$ (the diameter of the circle in this case).

To obtain the analytic lift-based adjoint solution, the streamline integral $\Xi(\vec{x}_0)$ has to be evaluated at each point of the domain. As with Giles and Pierce's analytic quasi-1D adjoint solution (Giles & Pierce, 2001), this cannot be done in analytic form in general, so the solution has to be evaluated numerically instead. In the present case, there is the additional complication that the integration requires prior determination of the streamline passing through $\vec{x}_0$. Thus, in order to produce sample solutions, a numerical procedure must be assembled to trace the streamline and perform the integration. Streamline tracing is carried out by direct numerical integration of the equation $d\vec{x}/dt = \vec{v}(\vec{x})$ with a fourth-order Runge-Kutta method (see e.g. (Giles & Haimes, 1990)). The integration is carried out concurrently with a simple trapezoidal rule. Contour maps of the analytic lift-adjoint solution obtained in this way are shown in Figure 6, while Figure 7 shows the adjoint solution along three vertical lines crossing the stagnation streamline upstream and downstream the circle and impinging the wall as indicated in Figure 6. It is clear from both figures that the solution is singular along the dividing streamline upstream of the rear stagnation point (the circle wall and the incoming stagnation streamline), but not across the rear stagnation streamline.

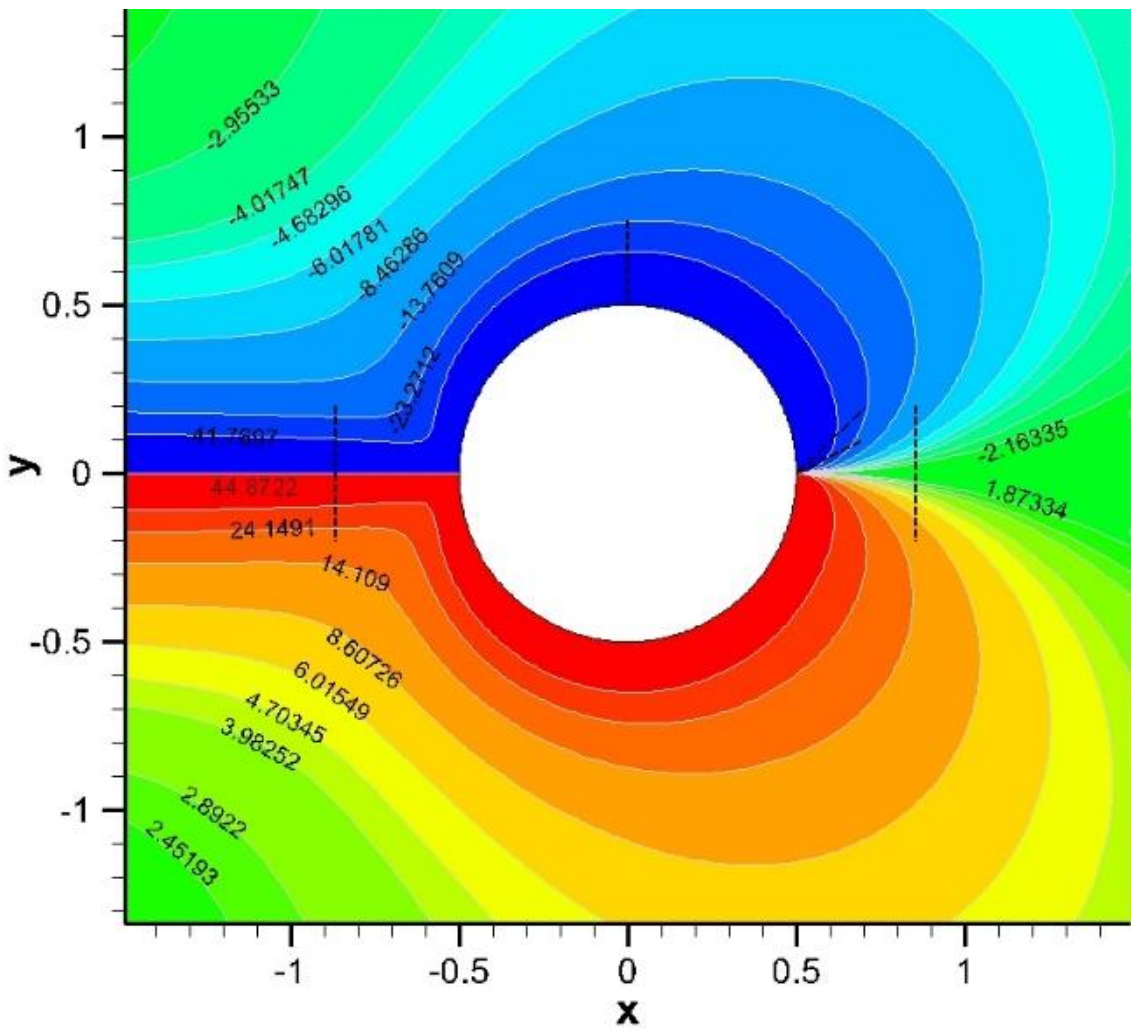

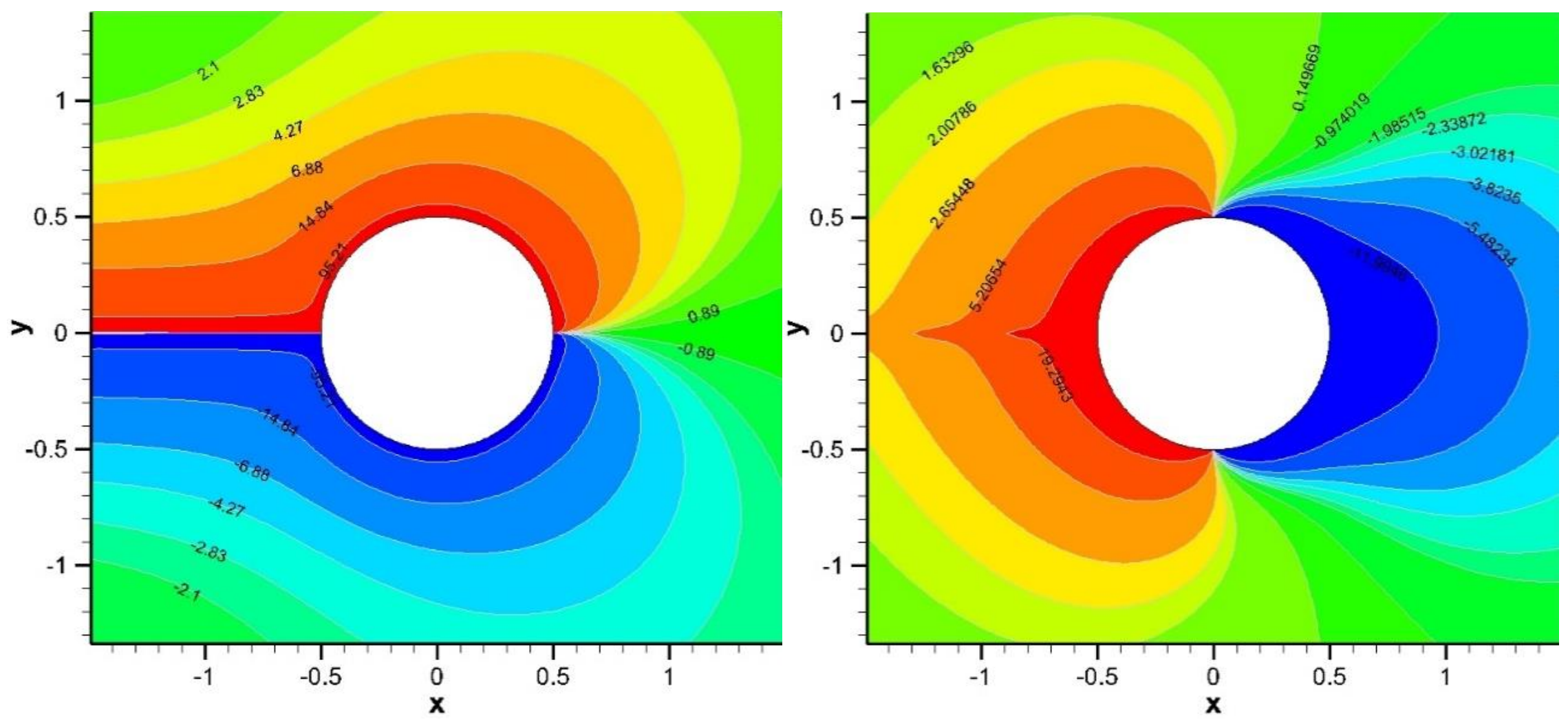

**Figure 6.** Analytic lift-based adjoint solution for incompressible, inviscid flow past a circle. Top: $\psi_1$. Bottom left: $\psi_2$. Bottom right: $\psi_3$.

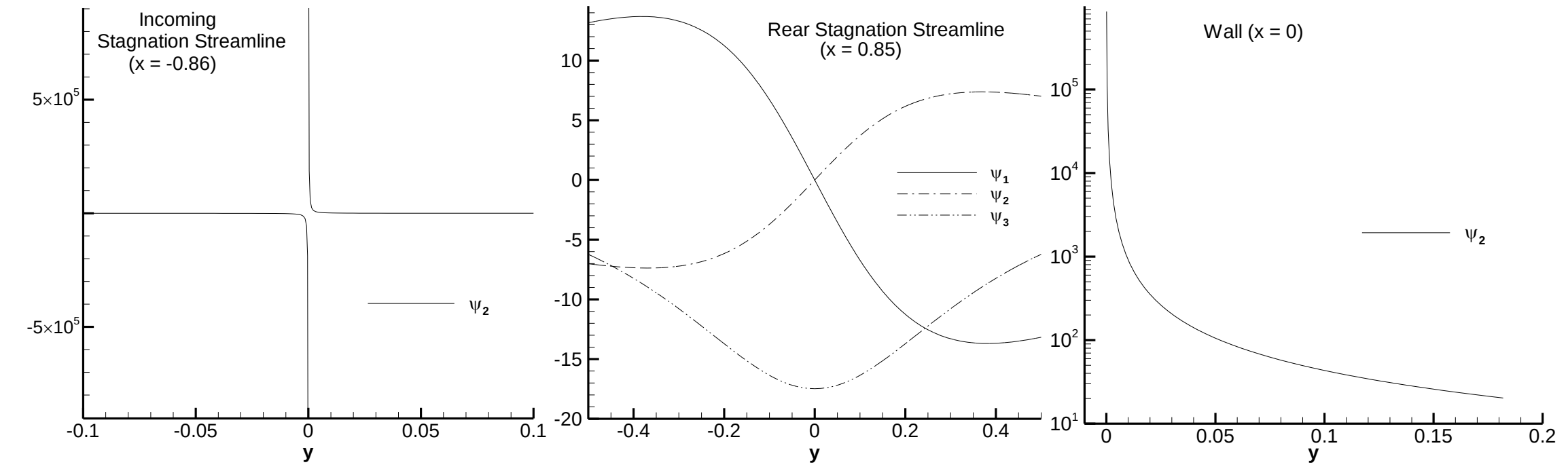

**Figure 7.** Analytic adjoint solution along vertical lines crossing the stagnation streamline upstream and downstream the circle and impinging the wall as indicated in Figure 6.

These singularities are caused by the streamline integral $\Xi$. In section 4 we argued that the streamline integral behaves as $\Xi \sim r_{\min}^{2-\frac{5}{k}} \sim r_{\min}^{-\frac{2\tau/\pi+1}{2-\tau/\pi}}$ in terms of the distance of the streamline to the rear stagnation point/trailing edge. Since $k = 1$ for the circle, we expect that $\Xi \sim r_{\min}^{-3}$ at the rear stagnation point and $\Xi \sim d^{-3/2}$ at the wall and the incoming stagnation streamline, which is confirmed in Figure 8, Figure 9 and Figure 10. Figure 8 plots the value of the integral $\Xi$ at several points placed along a vertical line crossing the incoming stagnation streamline (the $x$ axis) as indicated in Figure 6. This value is fitted against the distance $d$ to the stagnation streamline, yielding $\Xi \sim d^{-1.47}$, which is close to the theoretical prediction and to the value $\Xi \sim d^{-1.46}$ obtained at the wall (Figure 9), but also against the minimum distance of the streamline to the rear stagnation point $r_{\min}$, yielding $\Xi \sim r_{\min}^{-2.9}$, which is close to the theoretical value and to the value of the exponent obtained for the rear stagnation point (Figure 10). Finally, the integral vanishes linearly $\Xi \sim d$ across the rear stagnation streamline (Figure 11), which is not singular.

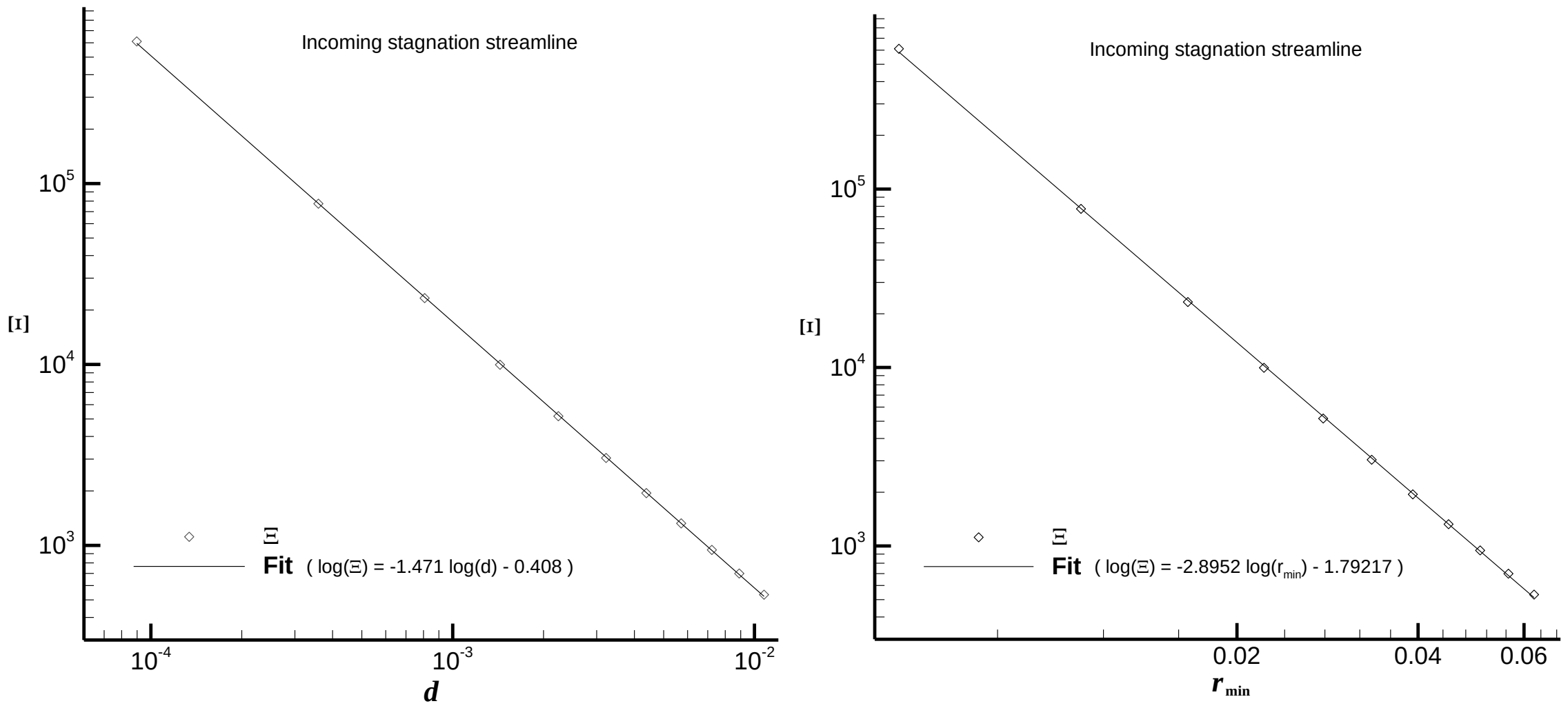

**Figure 8.** Value of the streamline integral computed with the analytic solution along a vertical line at $x = -0.86$ crossing the stagnation streamline upstream of the circle as indicated in Figure 6.

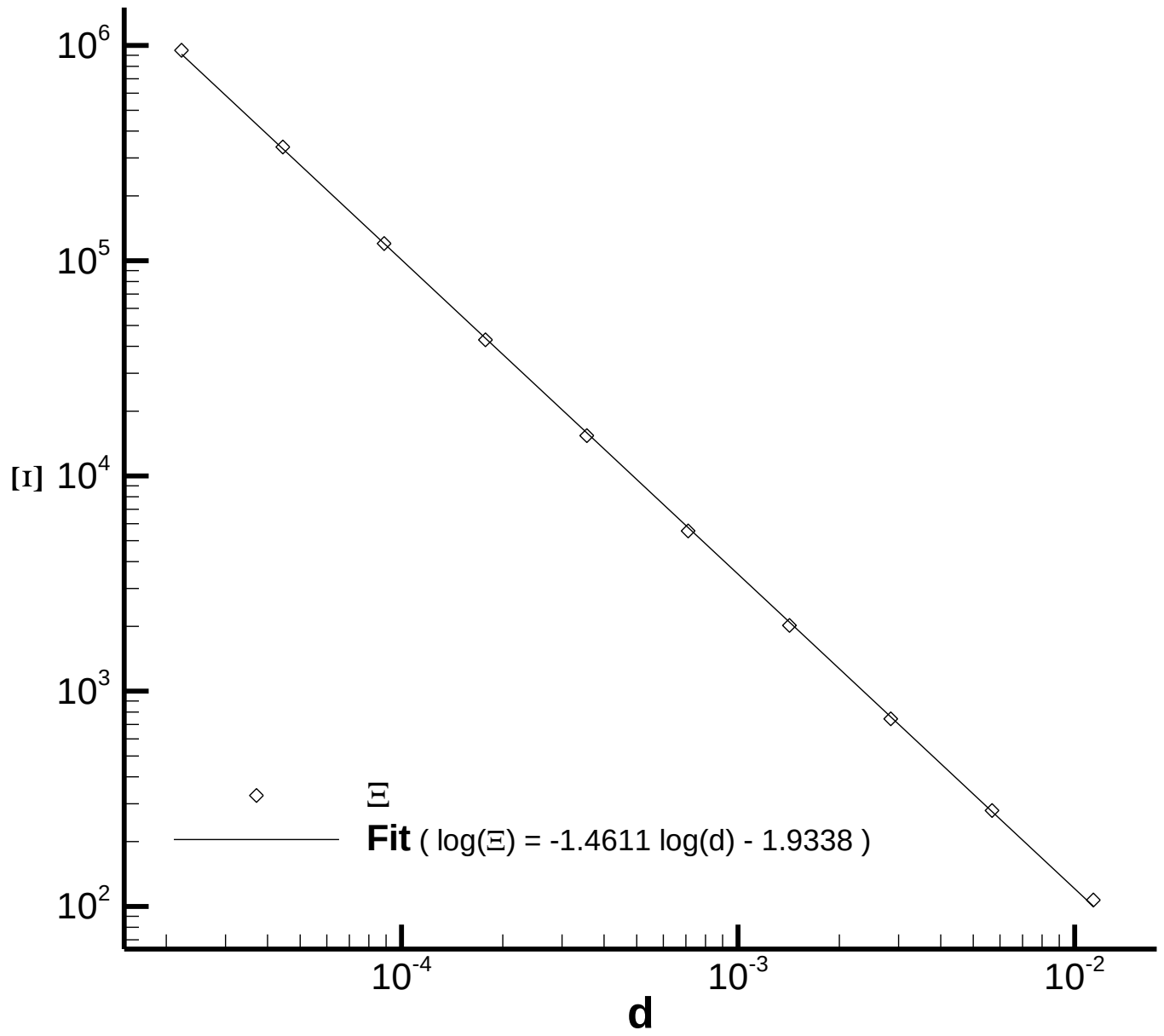

**Figure 9.** Value of the streamline integral computed with the analytic solution along the vertical line $x = 0$ impinging the circle wall as indicated in Figure 6.

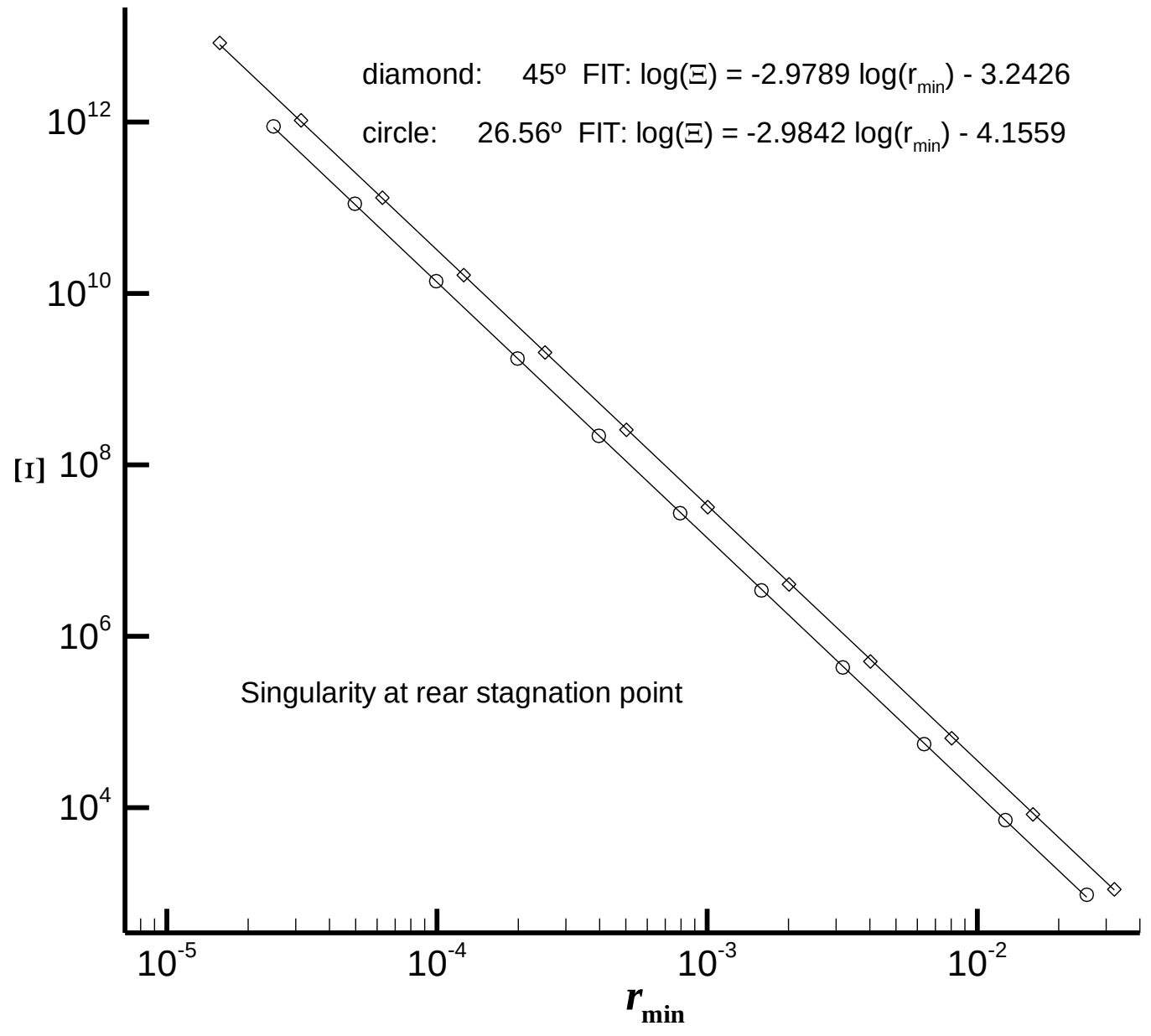

**Figure 10.** Value of the streamline integral computed with the analytic solution along two lines approaching the rear stagnation point with different inclinations as indicated in Figure 6.

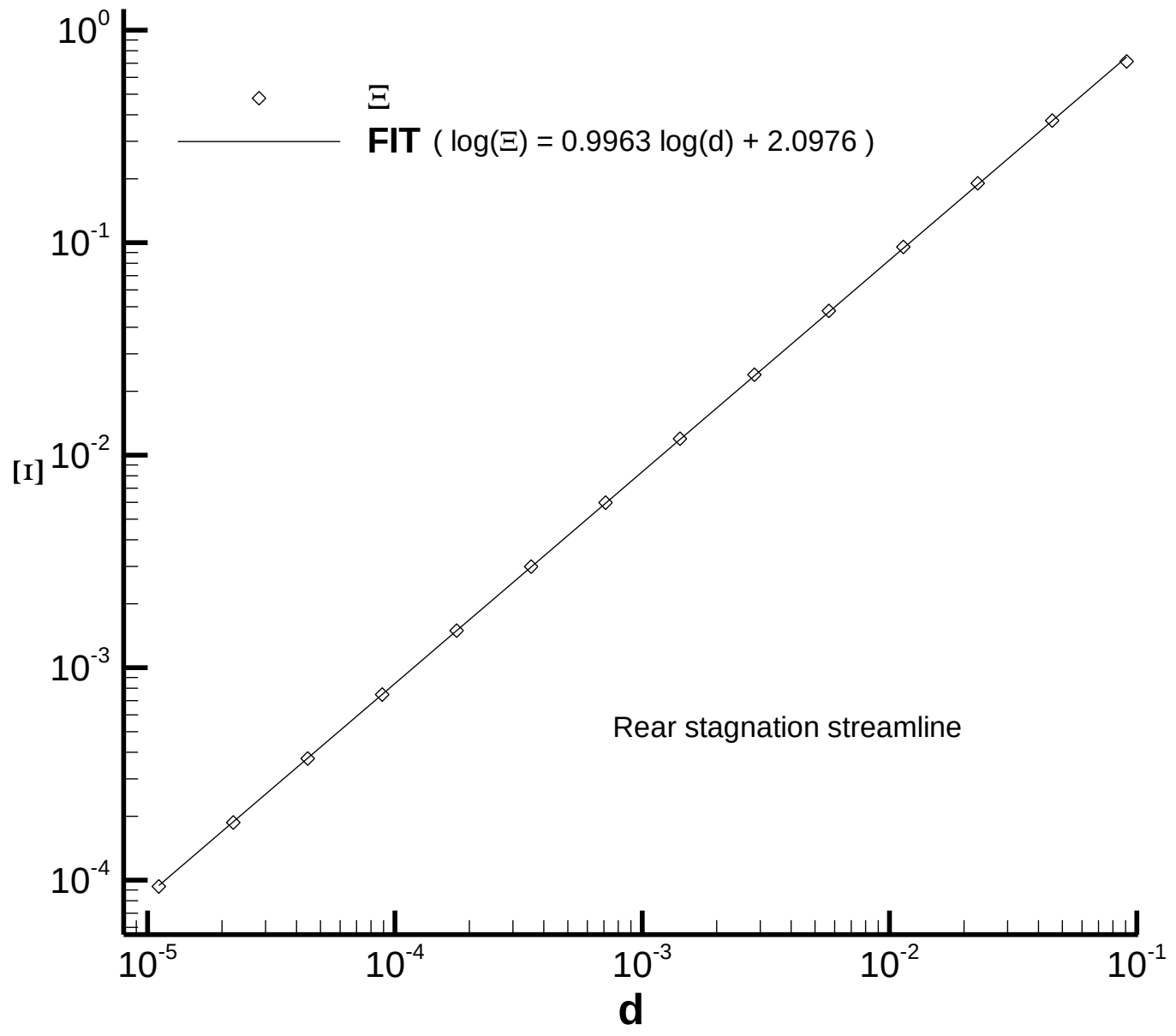

**Figure 11.** Value of the streamline integral computed with the analytic solution along a vertical line at $x = 0.85$ crossing the stagnation streamline downstream of the circle as indicated in Figure 6.

We have argued above that the behavior at the wall and stagnation streamline is essentially governed by the singularity exponent at the rear singularity, which is thus the primary singularity. The results in Figure 8, Figure 9 do support this idea, which is further illustrated with Figure 12 and Figure 13. Figure 12 (left) shows the value of $\Xi$ along the streamlines shown on the right plot. It can be clearly seen that the integral is quite approximately constant upstream of the trailing edge region and negligible downstream, and that the upstream value grows as the streamline approaches the dividing streamline (the downstream value, on the other hand, approaches zero), as expected. Figure 13, which focuses on the outermost streamline of Figure 12, plots the running value of Ξ along the streamline, as well as $I = -\partial_s q^{-2}\Upsilon^{(1)} + 2q^{-2}\partial_s\phi(1+\Upsilon^{(2)})$, which is the integrand in Ξ. It is clear that the main contribution to the integral comes from the region near the rear stagnation point, with a negligible contribution from the leading edge region.

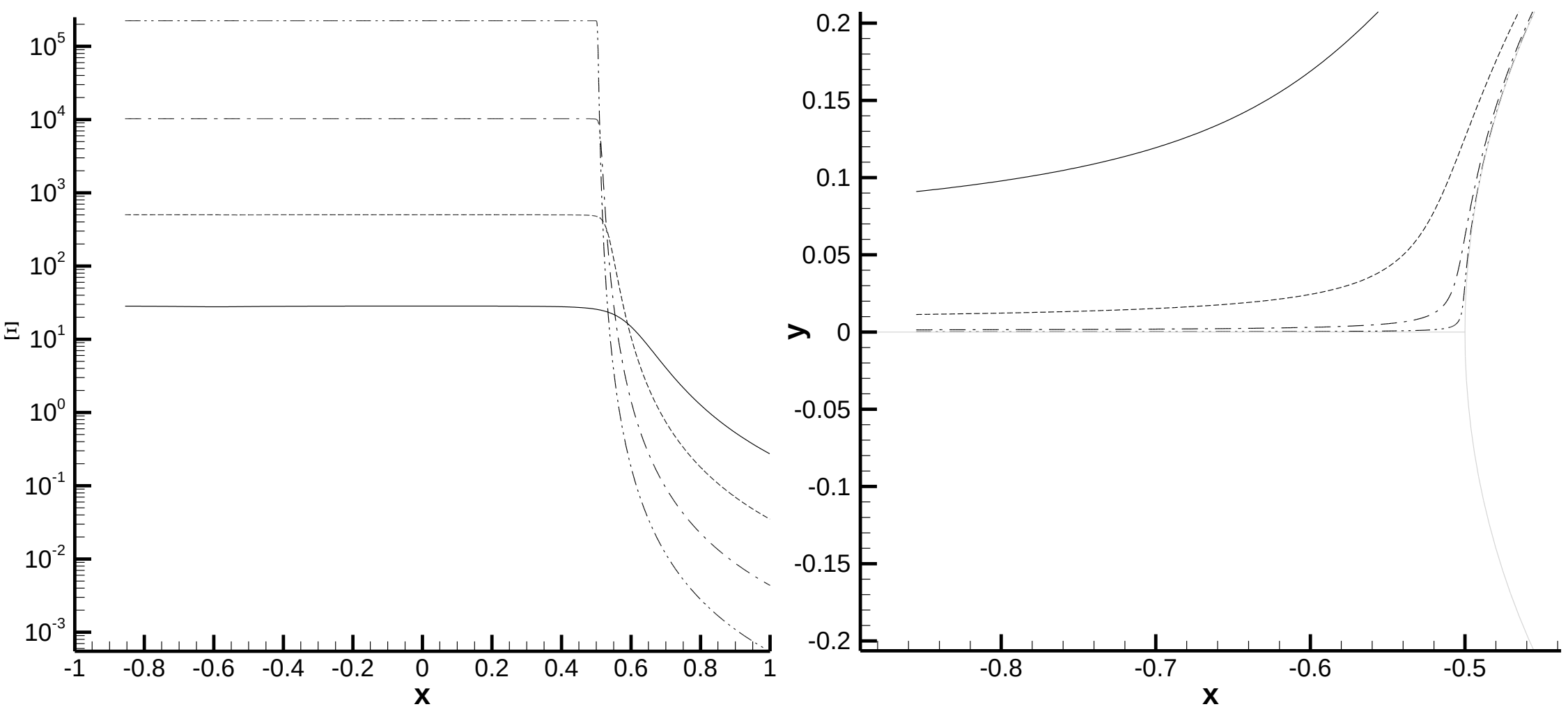

**Figure 12**. Left: running value of $\Xi(\vec{x}_0)$ along the streamlines indicated on the right plot. The forward and rear stagnation points are located at $x = -0.5$ and $x = 0.5$, respectively.

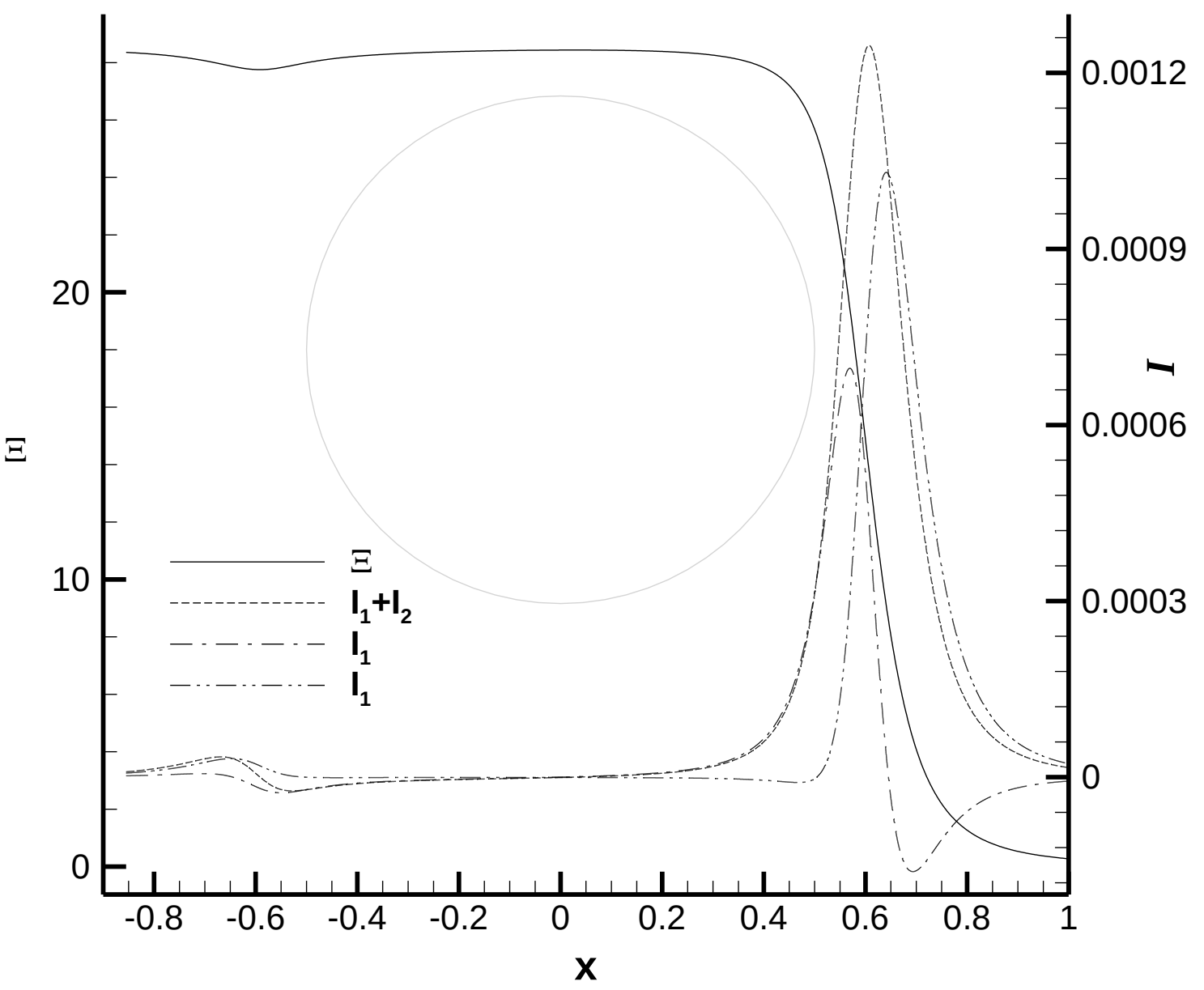

**Figure 13**. Running value of $\Xi(\vec{x})$ and integrands $I_1 = -\partial_s q^{-2}\Upsilon^{(1)}$, $I_2 = 2\partial_s\phi(1+\Upsilon^{(2)})/q^2$ and $I = I_1 + I_2$ along the outermost streamline of Figure 12. The extension of the circle is indicated for reference purposes.

Finally, Figure 14 plots the value of the linearized functionals $I^{(1)} = \psi_1 + \vec{v} \cdot (\psi_2, \psi_3)$ and $I^{(2)} = v\psi_2 - u\psi_3$ on the circle wall. These quantities have a finite value throughout the wall (except $I^{(1)}$, which diverges at the trailing edge), as expected.

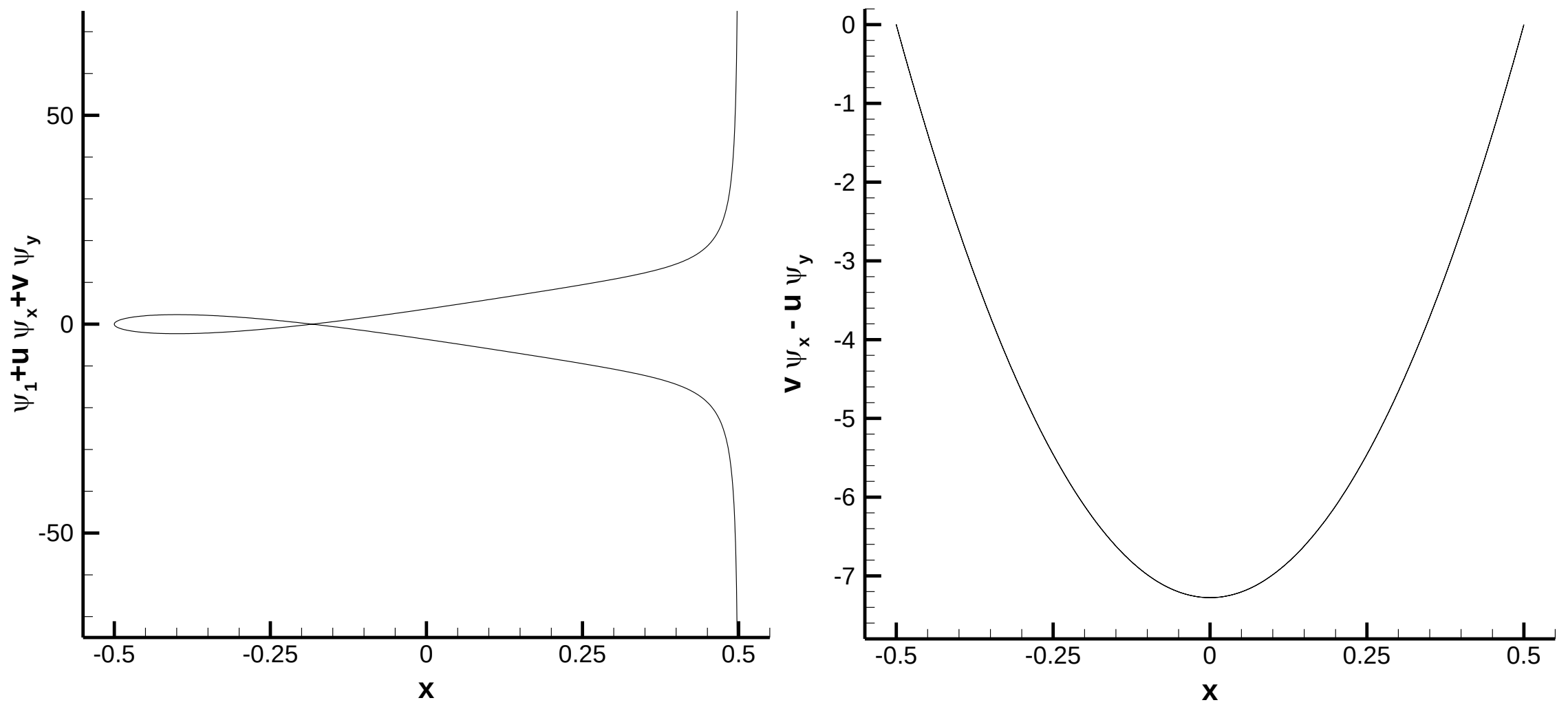

**Figure 14.** $I^{(1)} = \psi_1 + \vec{v} \cdot (\psi_2, \psi_3)$ and $I^{(2)} = v\psi_2 - u\psi_3$ at the wall computed with the analytic lift-based adjoint solution for incompressible, inviscid flow past a circle.

### 5.2. Airfoil with Finite Trailing-Edge Angle

The second case involves incompressible, inviscid flow at angle of attack $\alpha$ = 0° past a symmetrical van de Vooren airfoil given by the conformal transformation (Katz & Plotkin, 2001)

$$z = \frac{(\zeta - R)^k}{(\zeta - \varepsilon R)^{k-1}} + 1, \tag{69}$$

where $R = (1+\varepsilon)^{k-1} / 2^k$, $\varepsilon$ is a thickness parameter and $k$ is related to the trailing-edge angle $\tau$ as $k = 2 - \tau / \pi$. The transformation (69) maps the airfoil in the $z$ plane to a circle of radius $R$ centered at the origin in the $\zeta$ plane. In this paper, we set $\varepsilon = 0.0371$ and $k = 86/45$, resulting in an airfoil with 12% thickness and finite trailing edge angle $\tau = 16°$ that is shown in Figure 15, where the analytic flow solution obtained with conformal transformation techniques is also depicted. The analytic flow solution can be found in (Katz & Plotkin, 2001) and can be obtained from the solution in the circle plane (24) and the transformation (69) as $(u - iv)_{airfoil} = (d\Phi / d\zeta) / (dz / d\zeta)$.

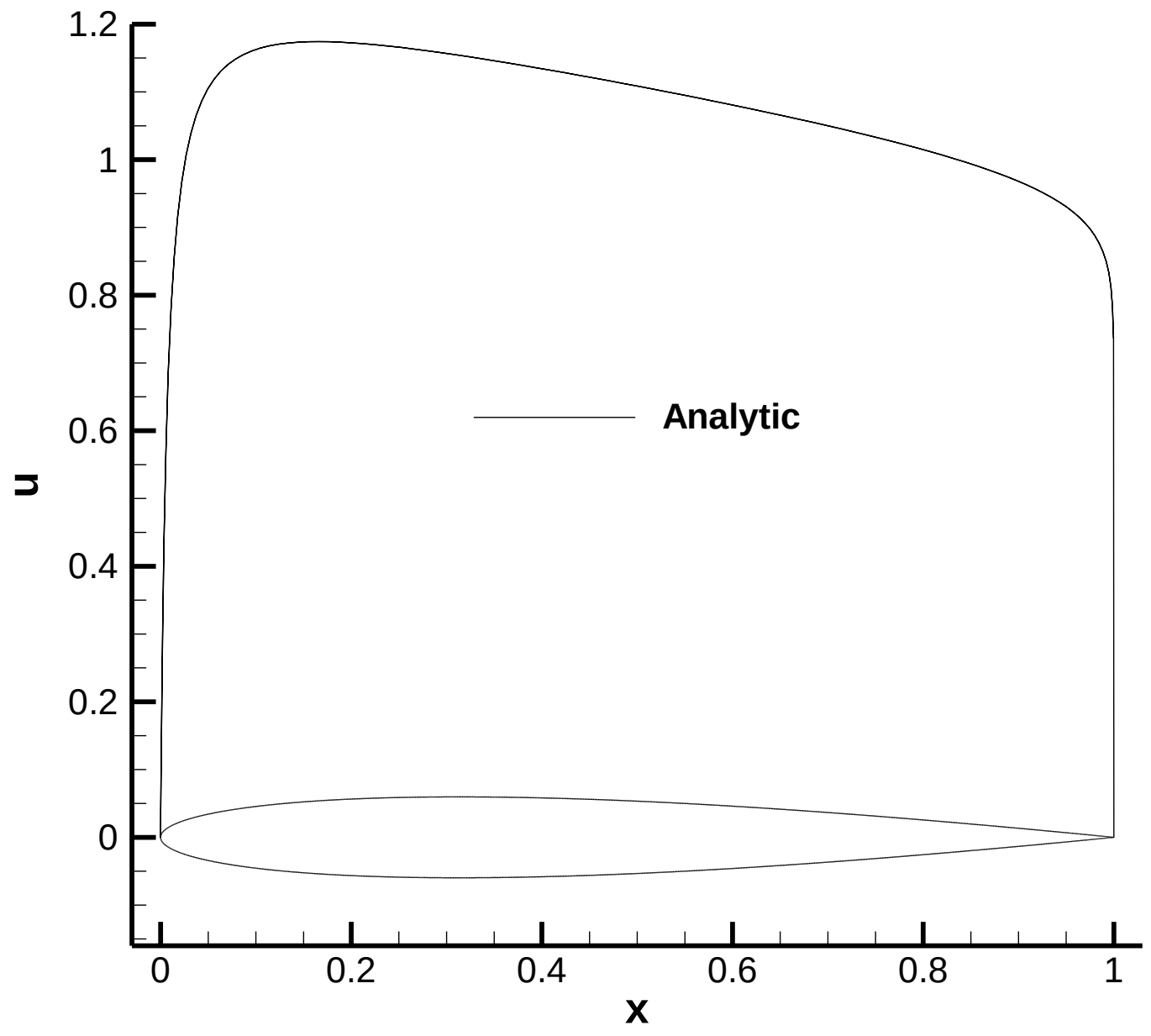

**Figure 15.** Geometry and analytic $x$-velocity for incompressible, inviscid flow at $\alpha$ = 0º past a van de Vooren airfoil with trailing-edge angle $\tau = 16°$ and 12% thickness computed with conformal transformation techniques. The flow is normalized such that $\rho_\infty = 1 = q_\infty$ and the airfoil profile is shown for reference.

Using this analytic flow solution, the corresponding analytic adjoint solutions can be computed using Eq. (56) and (60). Figure 16 shows the analytic drag-based adjoint solution. This solution has been compared to numerical solutions obtained by discretizing the adjoint equation directly in (Lozano & Ponsin, 2021) (Lozano & Ponsin, 2022).

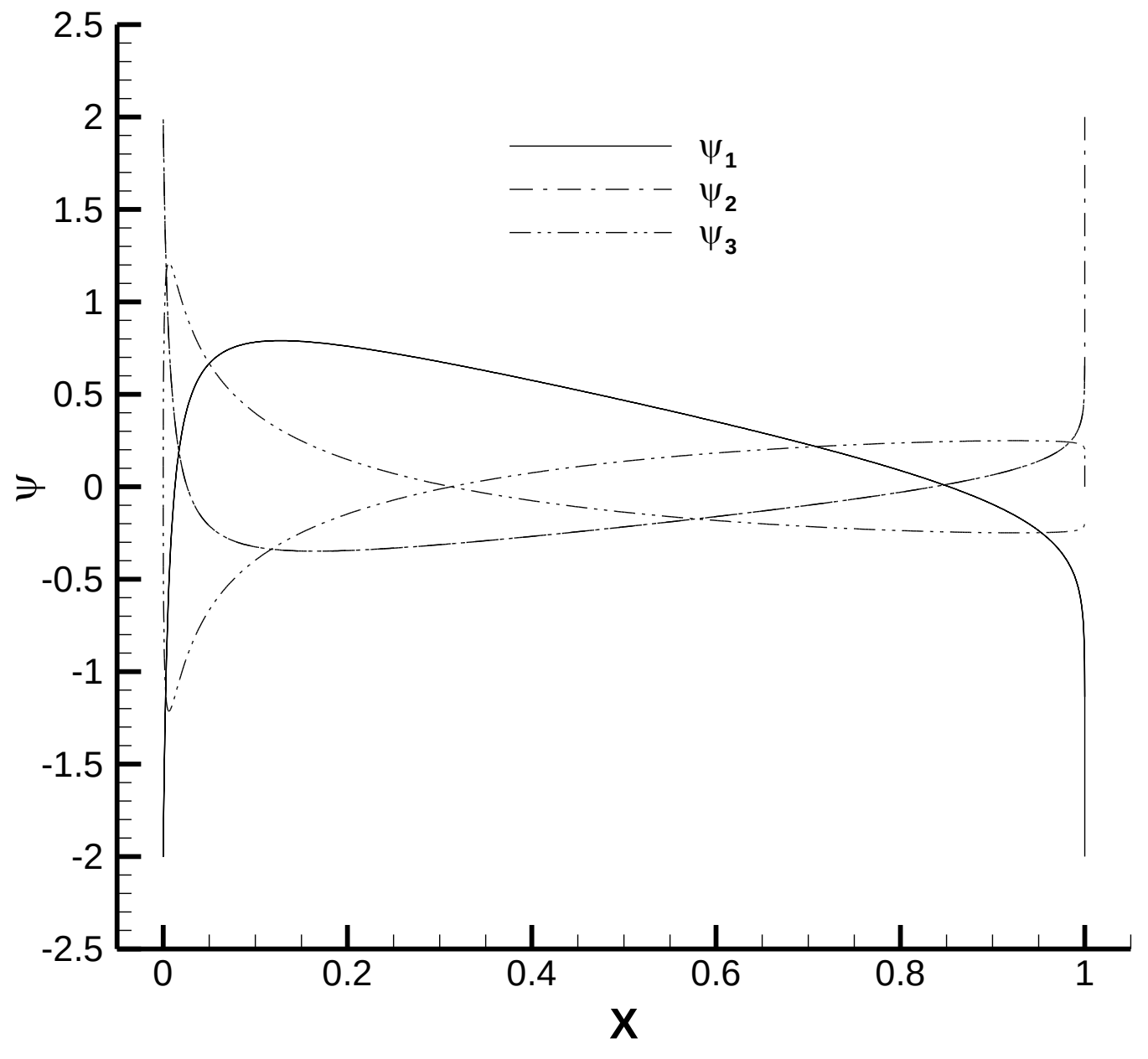

**Figure 16.** Analytic drag-based adjoint solution on the airfoil profile for incompressible, inviscid flow at $\alpha$ = 0º past a van de Vooren airfoil with trailing-edge angle $\tau = 16°$ and 12% thickness.

As for lift, the computation of the analytic adjoint solution is carried out with the fourth-order Runge-Kutta integrator described above. The streamline tracing is performed in the circle plane and then translated into the airfoil plane via the conformal transformation (69). Contour maps of the analytic lift adjoint solution obtained in this way are shown in Figure 17. Again, a comparison of this solution to numerical adjoint solutions can be found in (Lozano & Ponsin, 2022).

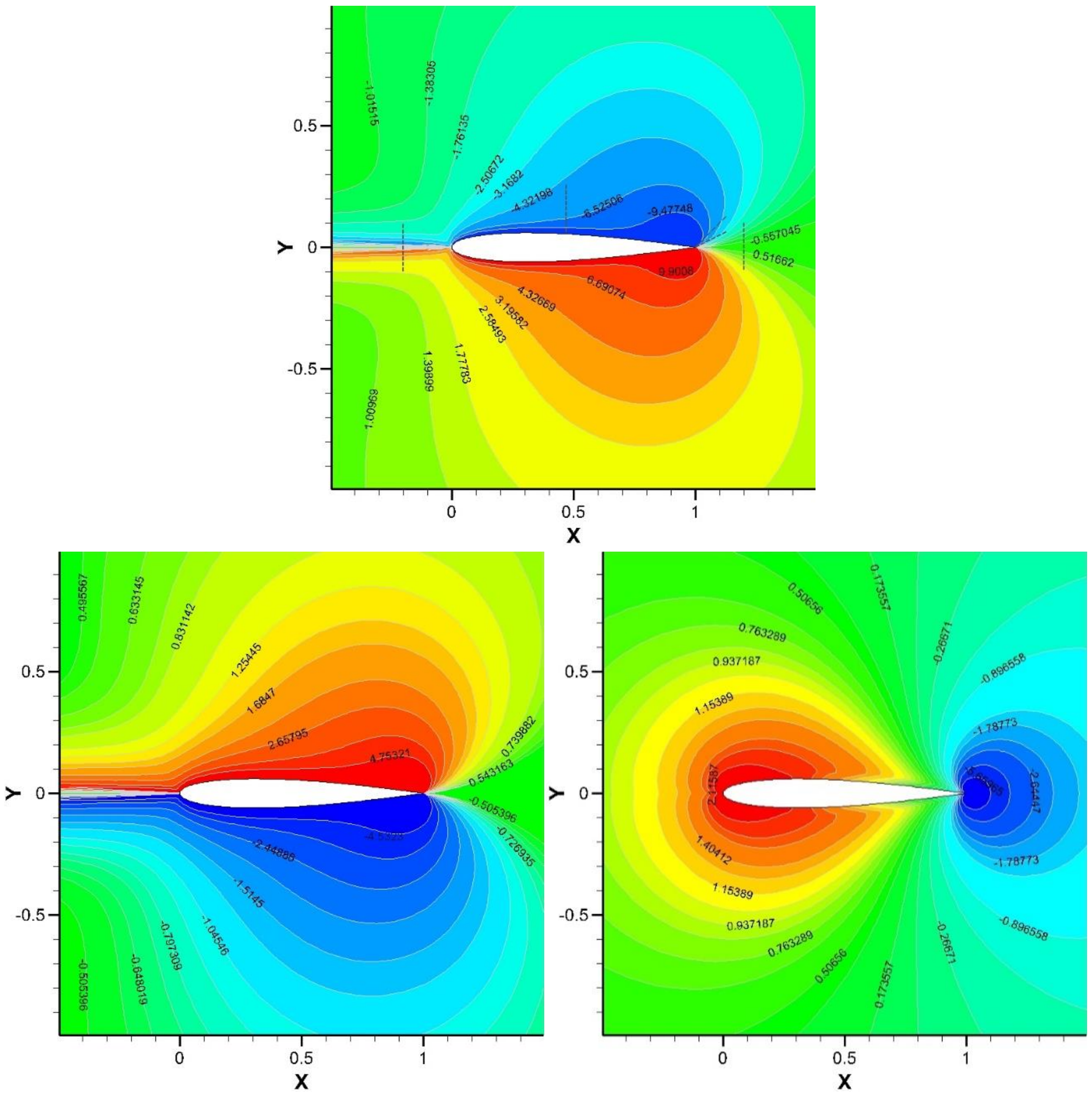

**Figure 17.** Analytic lift-based adjoint solution for incompressible, inviscid flow at $\alpha$ = 0º past a van de Vooren airfoil with trailing-edge angle $\tau = 16°$ and 12% thickness. Top: $\psi_1$. Bottom left: $\psi_2$. Bottom right: $\psi_3$.

As in the circle case, the analytic solution shows singularities at the wall, across the incoming stagnation streamline and at the trailing edge, but not across the rear stagnation streamline. This is more clearly illustrated in Figure 18, Figure 19, Figure 20 and Figure 21, which plot the streamline integral $\Xi$ along lines approaching the stagnation streamline upstream of the airfoil, the wall, the trailing edge and the rear stagnation streamline, respectively, as indicated in Figure 17. In all cases, the behavior of the integral has been illustrated by fitting the curves to a power of the distance to the dividing streamline or the trailing edge as appropriate. At both the incoming stagnation streamline and the wall, the integral diverges as roughly $1/d^{0.58}$, where $d$ is the distance to the stagnation streamline or to the wall. Noting that $k=86/45$ for the present case, the exponent is close to the theoretical prediction $\Xi \sim 1/d^{53/90} = 1/d^{0.589}$. At the trailing edge, the integral shows a $1/r^{0.61}$ singularity (where $r$ is the distance to the trailing edge) regardless of the inclination of the line, which is again in agreement with the theoretical prediction $\Xi \sim 1/r^{53/86} = 1/r^{0.616}$. Finally, the streamline integral vanishes as $\Xi \sim d$ at the stagnation streamline downstream of the trailing edge.

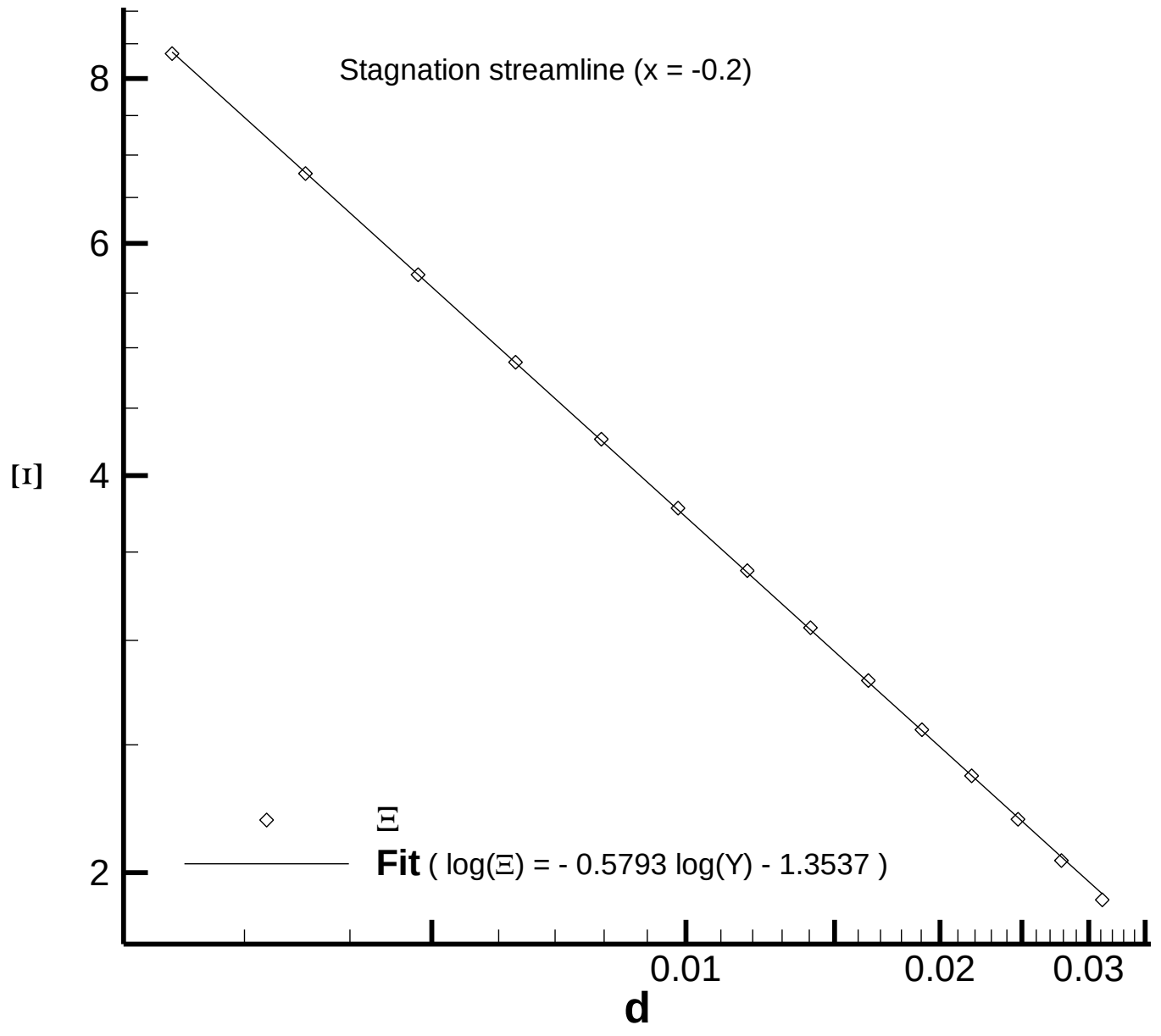

**Figure 18.** Value of the streamline integral computed with the analytic solution along a line crossing the stagnation streamline upstream of the airfoil as indicated in Figure 17.

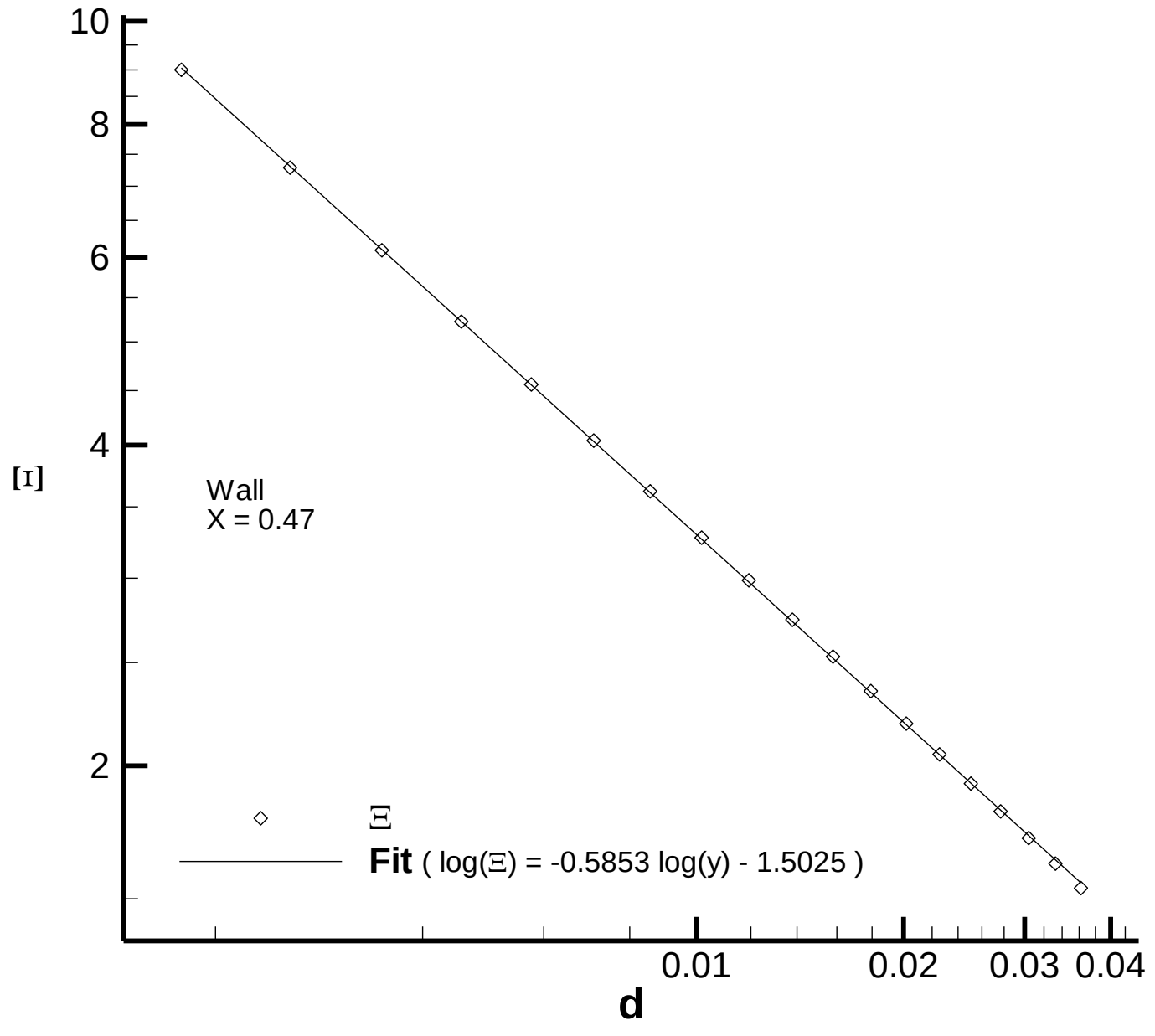

**Figure 19.** Value of the streamline integral computed with the analytic solution along a line impinging the airfoil as indicated in Figure 17.

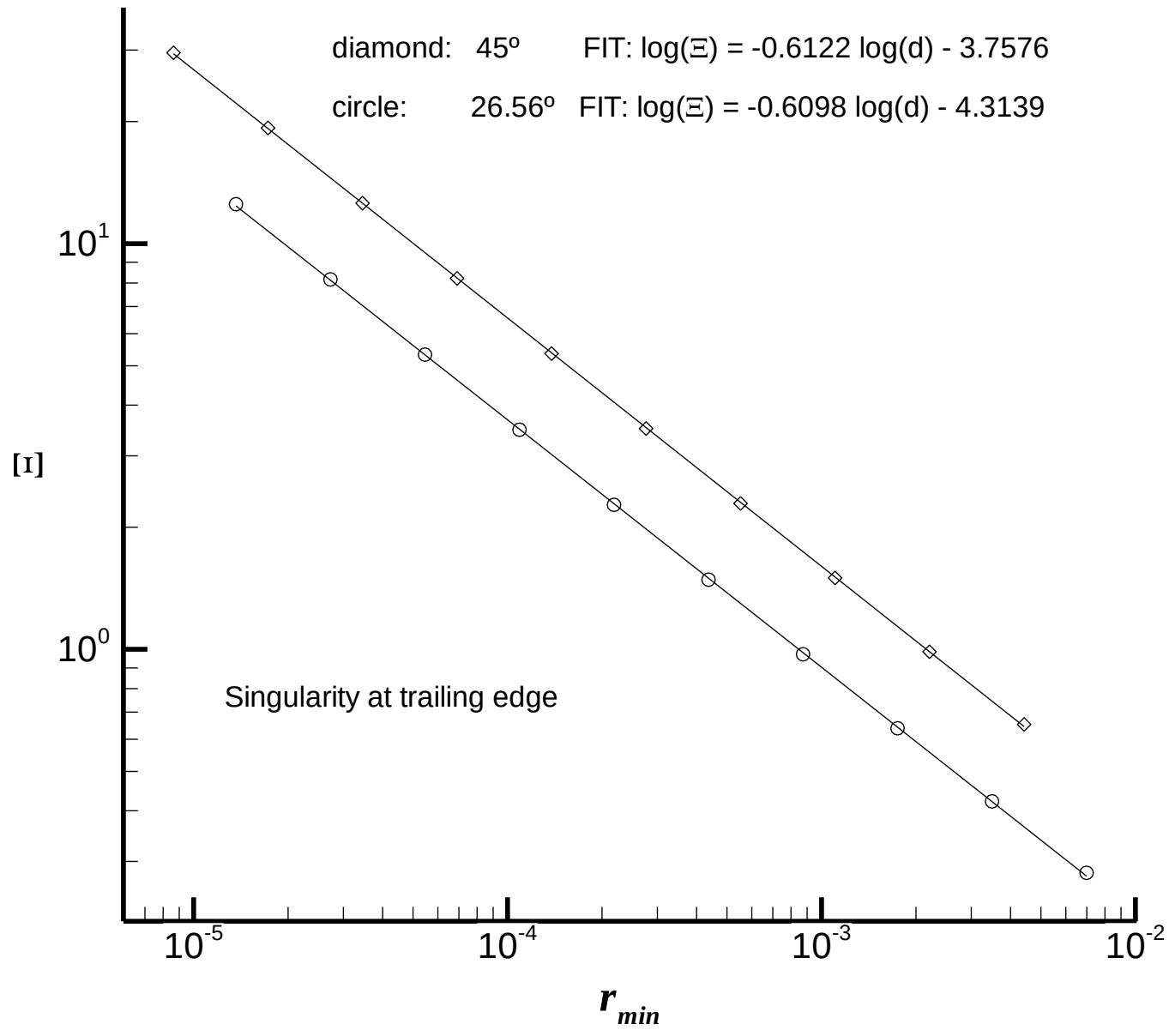

**Figure 20.** Value of the streamline integral computed with the analytic solution along two lines approaching the trailing edge at different inclinations as indicated in Figure 17.

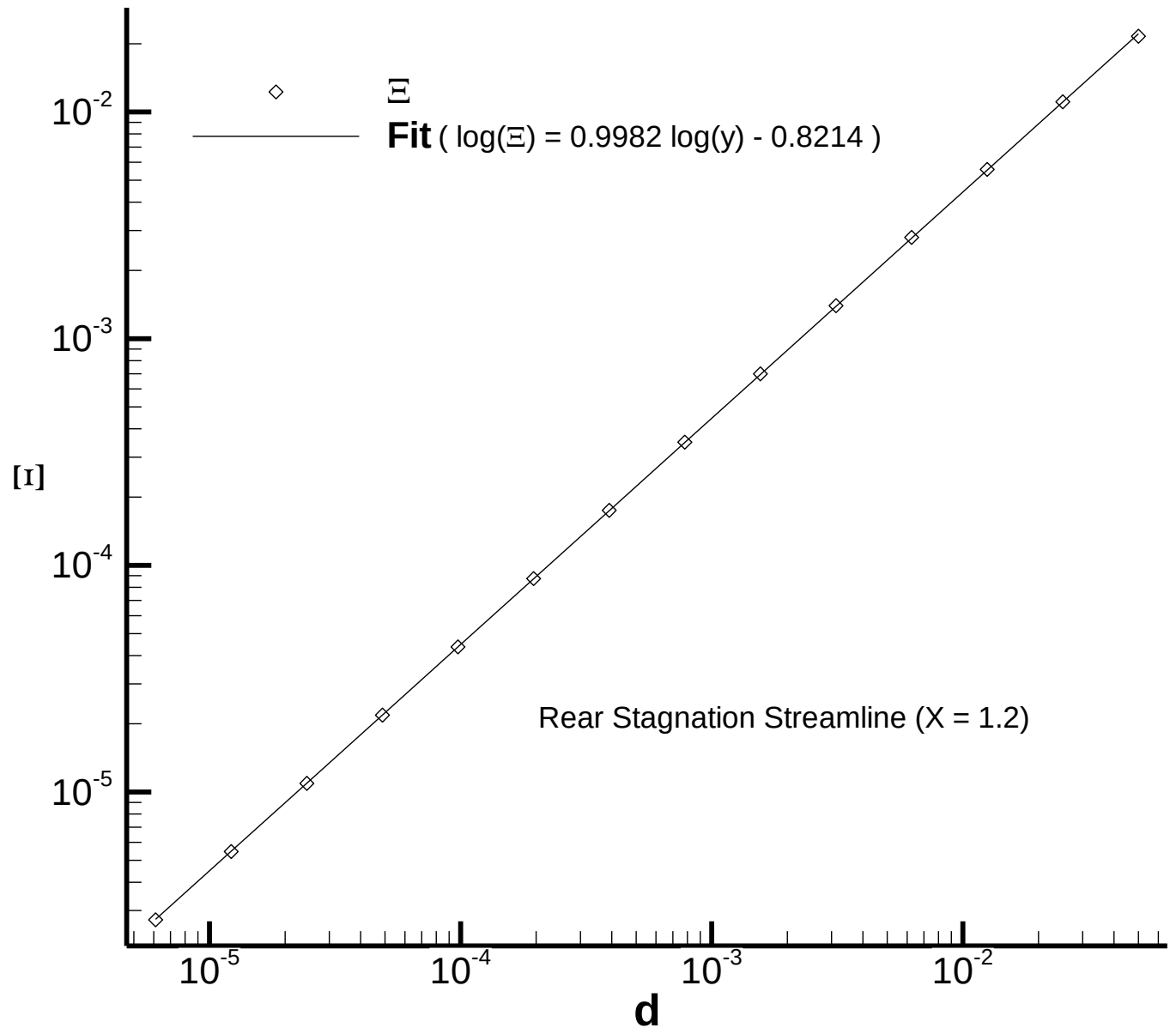

**Figure 21.** Value of the streamline integral computed with the analytic solution along a line crossing the stagnation streamline downstream of the airfoil as indicated in Figure 17.

Figure 22 shows the value of $I = -\partial_s q^{-2}\Upsilon^{(1)} + 2q^{-2}\partial_s\phi(1+\Upsilon^{(2)})$ along a streamline originating near the stagnation streamline upstream of the airfoil. It is clear, as in the previous case, that the main contribution to the integral comes from the region near the trailing edge, with a negligible contribution from the leading edge region.

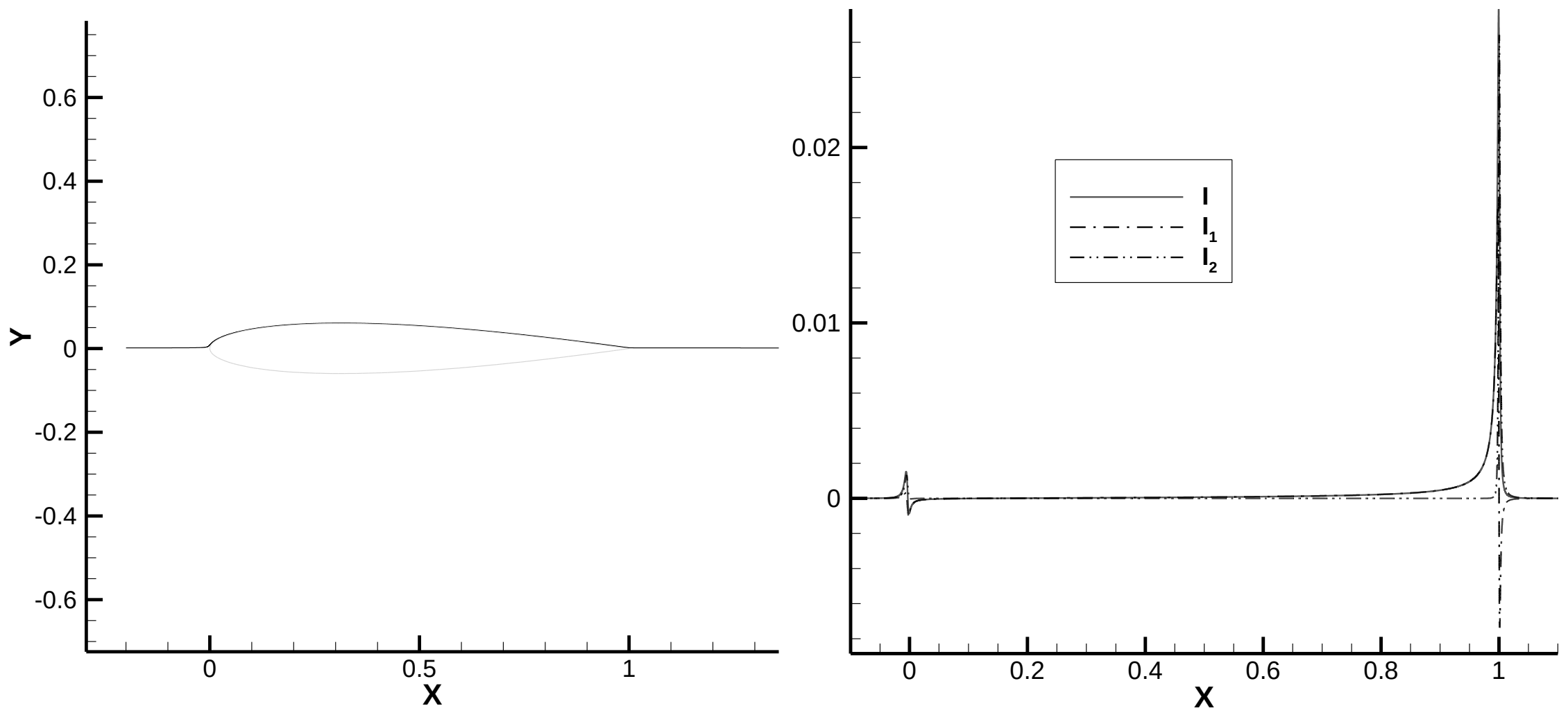

**Figure 22.** Plot of $I_1 = -\partial_s q^{-2}\Upsilon^{(1)}$, $I_2 = 2\partial_s \phi(1+\Upsilon^{(2)})/q^2$ and $I = I_1 + I_2$ for the streamline shown on the left. Inviscid, incompressible flow at $\alpha$ = 0º past a van de Vooren airfoil with trailing-edge angle $\tau = 16°$ and 12% thickness.

To end the analysis of the airfoil solution, Figure 23 focuses on the behavior of the adjoint solution near the airfoil profile. In the plot, the first component of the analytic adjoint solution is shown for a succession of O-shaped curves surrounding the van der Vooren airfoil profile and progressively closer to it, clearly demonstrating the divergence of the analytic solution at the wall. It is clear that the anomaly observed in numerical computations (Lozano, 2019) (Peter, et al., 2022) (Lozano & Ponsin, 2022) is caused by a divergence of the analytic solution at the wall. On the numerical side, the numerical viscosity of the solver stabilizes the divergence, producing a finite value at the profile which nevertheless varies continually as the grid spacing or the intensity of the numerical dissipation change (Lozano & Ponsin, 2021). At any rate, while the adjoint variables diverge at the wall, the linearized functionals $I^{(1)}$ and $I^{(2)}$ (66) are again finite across the profile (except $I^{(1)}$ which diverges at the trailing edge) as shown in Figure 24.

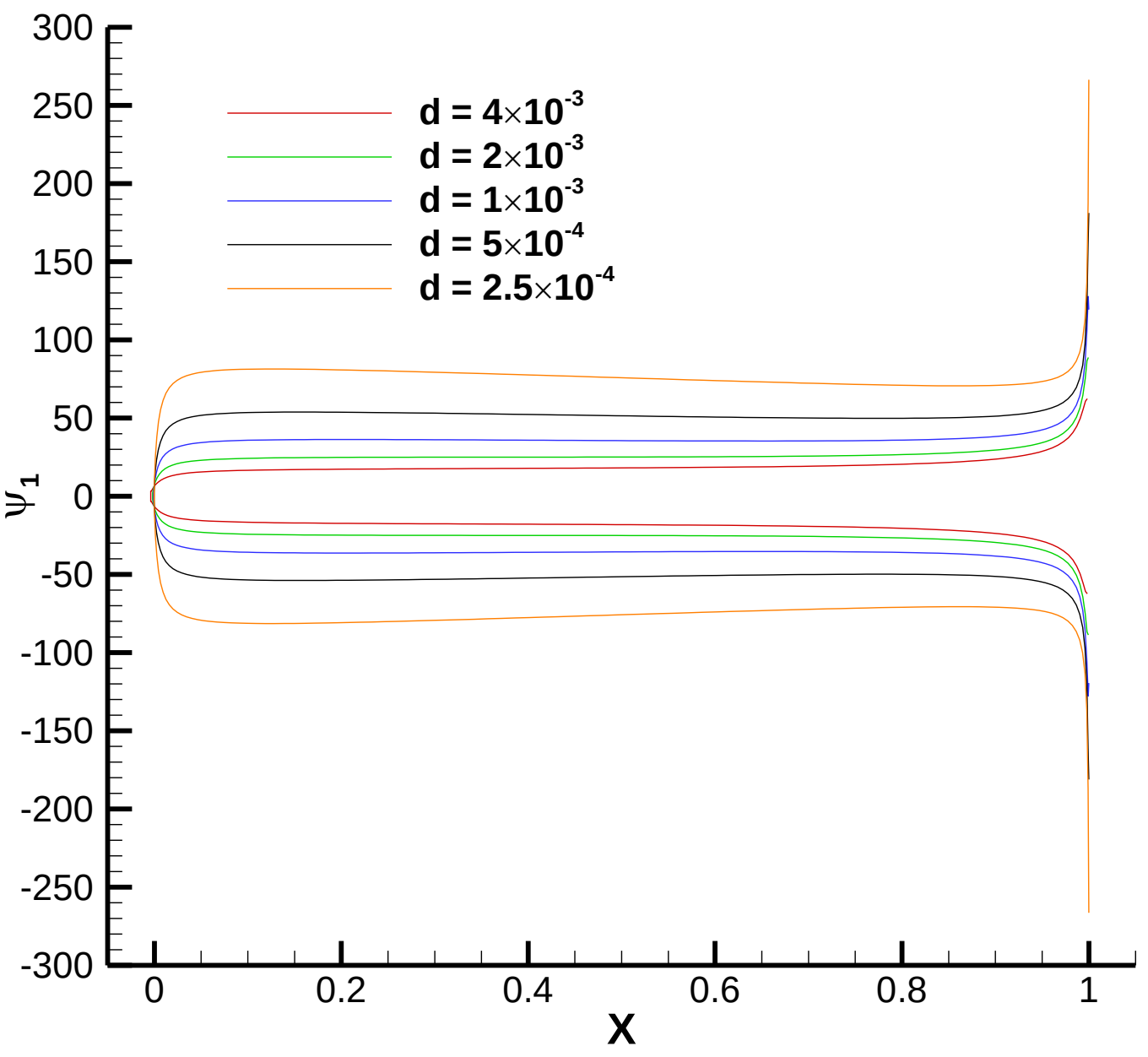

**Figure 23.** Analytic lift-based adjoint solution for incompressible, inviscid flow at $\alpha$ = 0º past a van de Vooren airfoil with trailing-edge angle $\tau = 16°$ and 12% thickness on a sequence of O-shaped curves progressively approaching the airfoil.

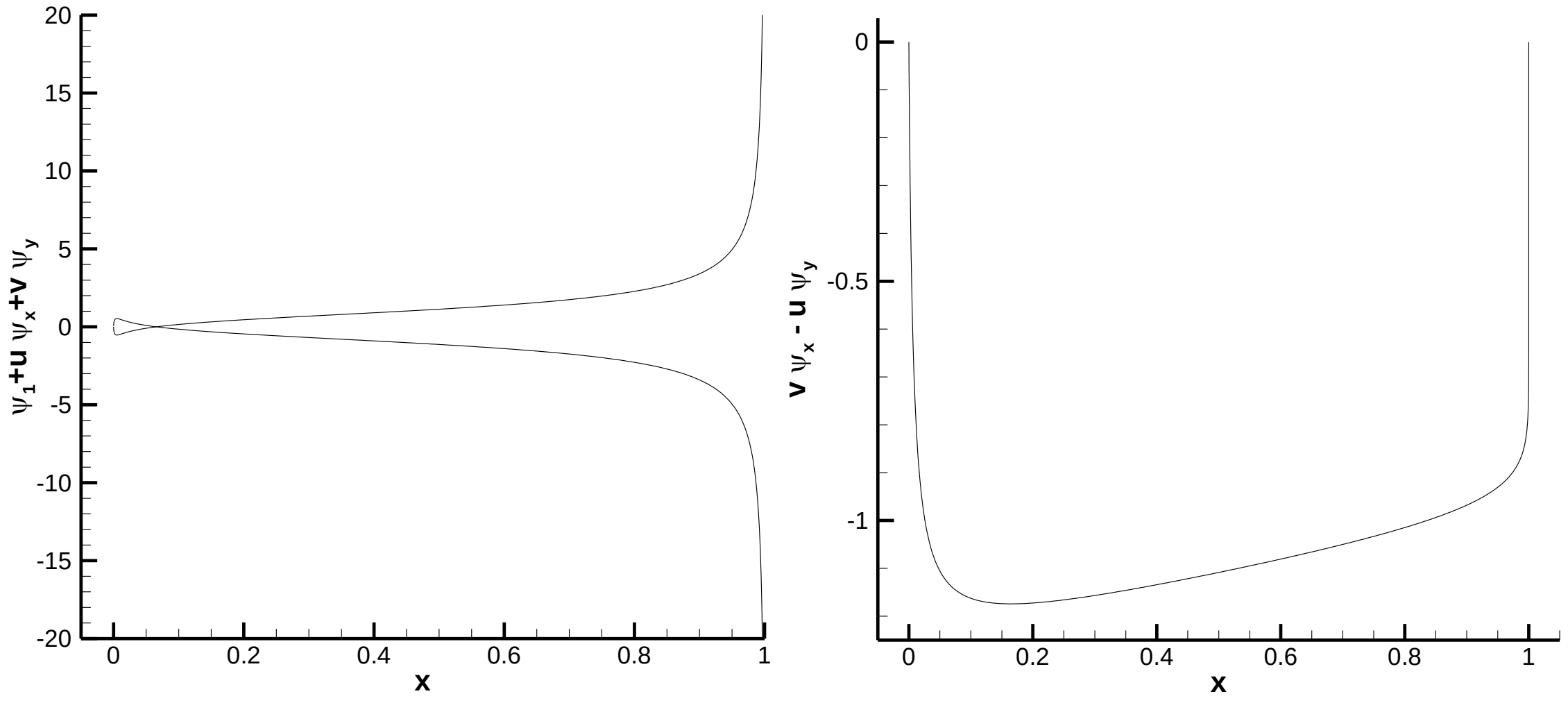

**Figure 24.** $I^{(1)} = \psi_1 + \vec{v} \cdot (\psi_2, \psi_3)$ and $I^{(2)} = v\psi_2 - u\psi_3$ on the airfoil profile computed with the analytic lift-based adjoint solution for incompressible, inviscid flow at $\alpha$ = 0º past a van de Vooren airfoil with trailing-edge angle $\tau = 16°$ and 12% thickness.

## 6. Conclusions

In this paper, we have investigated the adjoint solutions for the two-dimensional incompressible Euler equations. By restricting to irrotational base flows, it is possible to derive a closed form solution to the adjoint equations using the Green's function approach developed by Giles and Pierce (Giles & Pierce, 2001). Drag-based and lift-based analytic adjoint solutions have been obtained in this way, and they should be of help as test cases for verification of numerical adjoint solvers.

The drag-based adjoint solution is smooth and has a particularly simple closed-form expression in terms of the flow variables. For the type of flows considered in this paper, drag is zero for the analytic solution and has thus no practical application for shape design. The drag-based adjoint solution is, however, non-trivial and can be used for verification purposes. Likewise, if drag fails to vanish in a discrete flow solution, the drag-adjoint solution can be used to build a cheap mesh adaptation indicator that will identify those flow regions responsible for the failure, in analogous fashion to the entropy adjoint approach (Fidkowski & Roe, 2010).

The lift-based adjoint solution, on the other hand, is considerably more complex, and exhibits a number of distinctive features, some of which had been already anticipated in the literature (Giles & Pierce, 1997) but are now fully clarified. The lift adjoint solution shows a singularity at the trailing edge/rear stagnation point that corresponds to perturbations to the Kutta condition. The sensitivity of the Kutta condition to perturbations to the stagnation pressure gives rise to an additional singularity structure along the dividing streamline upstream of the rear singularity, resulting in adjoint singularities at the wall and the incoming stagnation streamline, where the adjoint solution diverges with identical exponents that nevertheless depend on the trailing edge angle. Since the analytic lift-adjoint solution is singular at walls, this finally settles the issue of the mesh divergence of inviscid adjoint solutions at solid walls (Lozano, 2019) (Peter, et al., 2022).

It would be interesting to extend this analysis to compressible flow. The task was initiated in (Giles & Pierce, 1997), where four linearly independent source terms were defined and a preliminary analysis was carried out. However, obtaining the exact adjoint solutions would require to compute the Green’s functions corresponding to these source terms or, at least, their effect on drag and lift. To the best of our knowledge, this is an open problem which appears very difficult to solve exactly, particularly for transonic flows. At any rate, preliminary analysis (Giles & Pierce, 1997) (Peter, et al., 2022) shows that adjoint variables are continuous at shocks, albeit with discontinuous derivatives, and at sonic lines, as long as they are not orthogonal to the flow. On the other hand, numerical compressible adjoint solutions do show singularities along the incoming stagnation streamline (Giles & Pierce, 1997) (Peter, et al., 2022), at the trailing edge/rear stagnation point (Giles & Pierce, 1997) (Lozano, 2018) (Lozano, 2019) (Peter, et al., 2022), and at walls (in the form of mesh-diverging adjoint solutions) (Lozano, 2019) (Peter, et al., 2022), which are the analog of the singularities found in incompressible flow, but also along certain supersonic characteristics (Lozano, 2018) (Peter, et al., 2022) without an analog in the incompressible case. The former singularities are likely governed by a mechanism similar to the one described in this paper (this has already been proven for the incoming stagnation streamline in (Giles & Pierce, 1997)), and probably with the same exponents, at least for subsonic flows.

Giles and Pierce's seminal work (Giles & Pierce, 2001) has just turned 20 years old. We would like to finish by quoting their closing remarks which are still as valid today as they were then: "An improved understanding of the behavior of adjoint solutions is necessary both to rigorously establish the theoretical basis for engineering optimal design methods and to illuminate the role of the adjoint solution in numerical error analysis. (…) Future developments along these lines will lead to great improvements in accuracy for key engineering quantities such as lift and drag." We couldn't agree more.

## Funding

The research described in this paper has been supported by INTA and the Ministry of Defence of Spain under grants Termofluidodinámica (IGB99001) and IDATEC (IGB21001).

## Declaration of Interests

The authors report no conflict of interest.